\documentclass[10pt,letterpaper]{article}
\usepackage[top = 2.5cm, bottom = 4cm, left = 2.5cm, right =2.5cm]{geometry}

\usepackage{amsmath,amssymb}

\usepackage{changepage}

\usepackage[utf8x]{inputenc}

\usepackage{textcomp,marvosym}

\usepackage{cite}

\usepackage{nameref,hyperref}

\usepackage[right]{lineno}

\usepackage{microtype}
\DisableLigatures[f]{encoding = *, family = * }

\usepackage[table]{xcolor}

\usepackage{array}

\usepackage{setspace} 

\usepackage[aboveskip=1pt,labelfont=bf,labelsep=period,justification=raggedright,singlelinecheck=off]{caption}

\usepackage{doi}
\usepackage{lastpage,fancyhdr,graphicx}
\usepackage{epstopdf}
\fancyheadoffset[L]{2.25in}
\fancyfootoffset[L]{2.25in}

\usepackage{mathtools}
\DeclarePairedDelimiter{\diagfences}{(}{)}
\newcommand{\diag}{\operatorname{diag}\diagfences}
\usepackage{subcaption,cleveref,tikz}
\renewcommand{\footrule}{}
\nolinenumbers
\def\justifying{%
  \rightskip=0pt
  \spaceskip=0pt
  \xspaceskip=0pt
  \relax
}
\justifying
\begin{document}

\title{\vspace{1cm}Inferring age-specific differences in susceptibility to and infectiousness upon SARS-CoV-2 infection based on Belgian social contact data\vspace{-1cm}}
\date{}

\maketitle

\begin{center}
Nicolas Franco \textsuperscript{1,2*},
Pietro Coletti\textsuperscript{1},
Lander Willem\textsuperscript{3},
Leonardo Angeli\textsuperscript{1},
Adrien Lajot\textsuperscript{4},
Steven Abrams\textsuperscript{1,5},
Philippe Beutels\textsuperscript{3},
Christel Faes\textsuperscript{1},
Niel Hens\textsuperscript{1,3}
\\[0.3cm]
\bigskip{\it 
\textbf{1} Data Science Institute, Interuniversity Institute of Biostatistics and statistical Bioinformatics (I-BioStat), UHasselt, Hasselt, Belgium
\\[0.3cm]

\textbf{2} Namur Institute for Complex Systems (naXys) and Department of Mathematics, University of Namur, Namur, Belgium
\\[0.3cm]

\textbf{3} Centre for Health Economic Research and Modelling Infectious Diseases (CHERMID), Vaccine \& Infectious Disease Institute, University of Antwerp, Antwerp, Belgium
\\[0.3cm]

\textbf{4} Department of Epidemiology and Public Health, Sciensano, Brussels, Belgium
\\[0.3cm]

\textbf{5} Global Health Institute (GHI), Family Medicine and Population Health, University of Antwerp, Antwerp, Belgium
\\

\bigskip}

* Corresponding author: nicolas.franco@uhasselt.be
\end{center}
\vspace{1cm}

\begin{abstract}
Several important aspects related to SARS-CoV-2 transmission are not well known due to a lack of appropriate data. However, mathematical and computational tools can be used to extract part of this information from the available data, like some hidden age-related characteristics. In this paper, we present a method to investigate age-specific differences in transmission parameters related to susceptibility to and infectiousness upon contracting SARS-CoV-2 infection. More specifically, we use panel-based social contact data from diary-based surveys conducted in Belgium combined with the next generation principle to infer the relative incidence and we compare this to real-life incidence data. Comparing these two allows for the estimation of age-specific transmission parameters. Our analysis implies the susceptibility in children to be around half of the susceptibility in adults, and even lower for very young children (preschooler). However, the probability of adults and the elderly to contract the infection is decreasing throughout the vaccination campaign, thereby modifying the picture over time.
\end{abstract}
\thispagestyle{empty}
\clearpage

\section*{Introduction}
Since the start of the COVID-19 pandemic, a new respiratory disease caused by the SARS-CoV-2 coronavirus, many mathematical and statistical approaches have been considered to identify transmission dynamics and characteristics of the virus. Some of those characteristics are still not completely known due to the lack of appropriate data. However, these characteristics are necessary in order to correctly inform public health policies as well as to develop more advanced scientific tools like mathematical and computational models. Concerning COVID-19, as for most infectious diseases, it quickly became apparent that some of the disease characteristics are strongly age-dependent~\cite{Davies:2020wt}. In particular, the susceptibility to SARS-CoV-2 infection as well as the infectiousness upon infection may be lower for children than for adults and the elderly, as shown by many studies mostly based on statistical approaches on incidence data \cite{Wu:2020tk,Goldstein:2020ui,10.1001/jamapediatrics.2020.4573,Hu:2021ur,doi:10.1126/science.abe2424}. Knowledge of such a difference could have an important impact on public health strategies in terms of prioritization of vaccination or the choice of targeted non-pharmaceutical interventions.

In this study, we propose a different method to estimate heterogeneous transmission parameters related to relative susceptibility and infectiousness using derived information from social contact data, and we illustrate this method using Belgian data. Social contact surveys \cite{Hoang:2019uh} coupled with the next generation principle \cite{diekmann2000mathematical,SANTERMANS201514} have been used for years to estimate key epidemiological parameters such as the basic (and effective) reproduction number (i.e., the average number of new infections caused by a \textit{typical} infected individual during their entire infectious period in a (fully) susceptible population), relative incidence or differences in susceptibility \cite{FLASCHE2011125}. The first large-scale social contact study, POLYMOD~\cite{POLYMOD}, collected social contact patterns for eight European countries between May 2005 and September 2006. In 2020-2021, social contact data has been collected in the so-called CoMix survey~\cite{Coletti2020,Verelst2021,Jarvis:2020,Gimma2021}, initially in the United Kingdom, The Netherlands and Belgium and afterwards extended to other European countries. Comix collected timely social contact information during the COVID-19 pandemic.

Social contact data can be used as a proxy to model SARS-CoV-2 transmission using the so-called \emph{social contact hypothesis} \cite{Wallinga2006}, which implies that the age-specific number of infectious contacts is proportional to the self-reported age-specific number of social contacts by a proportionality factor. This proportionality factor, often denoted by $q$, assumes that the probability of transmission is homogeneous across the different age classes. In the current paper, we aim to disentangle and quantify the heterogeneous components of this proportionality factor further \cite{Goeyvaerts}, elucidating information on relative age-specific susceptibility and infectiousness. Our approach is based on the method used by \cite{FLASCHE2011125} to estimate susceptibility profiles for influenza A/H1N1. However, we have refined the method to include a larger number of age categories and applied the methodology to SARS-CoV-2 transmission using a numerical approach. These estimates could serve to inform heterogeneous COVID-19 mathematical models relying on social contact data, such as e.g. mechanistic models \cite{Willem:2021tl,ABRAMS2021100449,Franco2020.09.07.20190108,Coletti2021}. Social contact data are also used in \cite{Munday:2021wd,Chin} to derive heterogeneous contributions to SARS-CoV-2 transmission using an approach based on the reproduction number. We go one step further using an approach based on the relative incidence derived from the next generation principle.

More specifically, we use the CoMix social contact data combined with daily incidence data on the number of new confirmed COVID-19 cases in Belgium over the period December 2020 to May 2021 to estimate the proportionality factor and its heterogeneous unmeasured components. We disentangle potential sources of heterogeneity in the acquisition of SARS-CoV-2 infection especially focusing on the comparison between children (infant, primary and secondary school) and adults. We also estimate the time evolution of the transmission parameters for different adult age classes throughout the vaccination campaign as carried out in Belgium showing an evolution of the proportionality factors over time. Then we present an illustration of the utility of heterogeneous proportionality factors by comparing the reproduction number estimated from the CoMix social contact data to the ones estimated from incidence of cases and hospitalizations, respectively.

\section*{Results}

\subsection*{Estimation of susceptibility and infectiousness through proportionality factors}

The proportionality factor is assumed to be age-specific and denoted by $q_{ij}$ where $i$ and $j$ belong to some age classes. $q_{ij}$ could be further split into heterogeneous components:
\[  q_{ij} =  \tilde q\, a_i \, h_j,  \]
where:
\begin{itemize}
\item The vector $(a_i)$ represents age-specific differences in factors influencing transmission which are specifically related to the susceptibility of individuals, including, but not limited to, direct (immunological) susceptibility to infection upon exposure (e.g., due to age-specific heterogeneous risk behavior and/or compliance to non-pharmaceutical interventions not already captured by contact frequency, natural susceptibility from previous infection, differences in vaccination status, etc.). In order to distinguish it from direct susceptibility to infection, this vector will be referred to as \emph{$q$-susceptibility}.
\item The vector $(h_j)$ describes age-specific differences in factors influencing transmission which are specifically related to the infectiousness of  individuals, including, but not limited to, infectiousness after acquiring infection (e.g., due to differences in viral load upon exposure, proportion of asymptomatic individuals, differences in vaccination status, mask wearing, etc.). In order to distinguish it from infectiousness upon infection, this vector will be referred to as \emph{$q$-infectiousness}.
\item The remaining global proportionality factor $\tilde q$ captures any remaining residual effect and is of no relevance when considering relative q-susceptibility or q-infectivity.
\end{itemize}
In the remainder of this paper, we will talk about susceptibility or infectiousness when considering immunological aspects of disease transmission, while we will add the prefix $q$ whenever quantities can carry additional effects related to susceptible and infectious individuals in order to avoid any ambiguity. 

If we denote by $w = (w_j)$ a vector representing the relative incidence within age class $j$ (usually normalized such that $\sum_j w_j = 1$) and by $\mathbf{M}^T$ a matrix containing the social contact data (whose components $m_{ji}$ represent the average daily number of individuals of age $j$ who have a contact with a single individual of age $i$), then we have the following system:
\[ \tilde q\, \diag{a_i} \,\mathbf{M}^T\diag{h_j} \, w = R_t \,w, \]
where $R_t$ represents the reproduction number. The core matrix of this system, $\mathbf{K} =\tilde q\, \diag{a_i} \,\mathbf{M}^T\diag{h_j}$ is called the \emph{next generation matrix} and gives the number of new infections in a successive generation. Details concerning the construction of this matrix and the next generation principle can be found in Section \nameref{MandM}.

Using our method, we compare the social contact matrix $\mathbf{M}^T$ extracted from the CoMix social contact survey in Belgium \cite{Coletti2020} and its derived relative incidence $w$ to the incidence obtained from real-life data in Belgium coming from PCR positive tests \cite{Sciensano}. This method allows for an estimation of either the relative $q$-susceptibility $(a_i)$ or relative $q$-infectiousness $(h_j)$ by age class, while assuming that the other set of parameters is known from the literature (i.e., holding one of the two vectors fixed). The chosen age groups are $[0,6)$ years, $[6,12)$ years, $[12,18)$ years, $[18,30)$ years and subsequent 10-year age classes up to $80+$ years in order to account for the Belgian educational system. Due to the method, obtained results only have a relative interpretation, hence we present them under the assumption of a mean susceptibility of one for the first adult class $[18,30)$. The period of observation goes from 22 December 2020 to 26 May 2021 (and to 15 June 2021 for confirmed cases data). A detailed description of considered data, literature assumptions, fitting procedure and normalization method is presented in Section~\nameref{MandM}.
\vspace{1cm}

\begin{figure}[h!]
\centering
\includegraphics[width=0.7\textwidth]{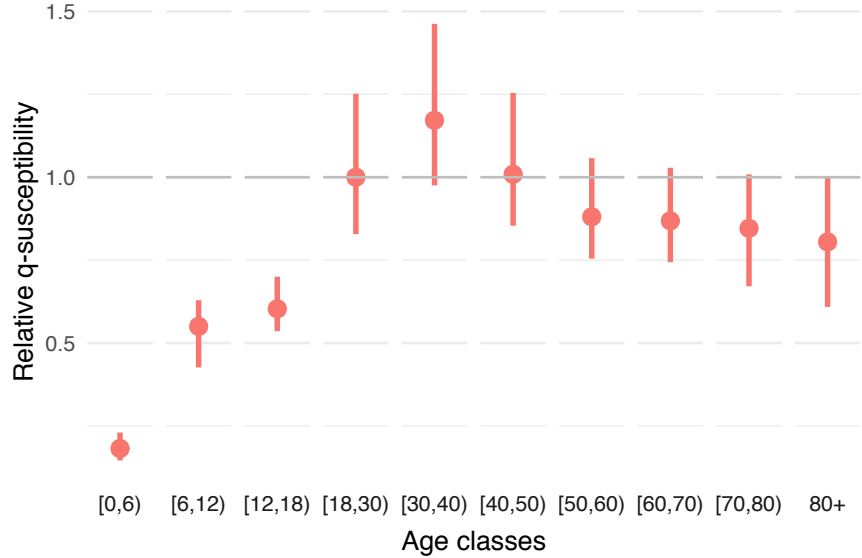}  
\caption{{\bf Estimated relative $q$-susceptibility.} The estimation of $q$-susceptibility is performed by age class using the next generation principle under an assumption on age-specific infectiousness $(0.54,0.55,0.56,0.59,0.7,0.76,0.9,0.99,0.99,0.99)$. The calibration is performed on CoMix waves 12 to 23 (observation period: 22 December 2020 to 15 June 2021). Dots represent means and bars represent 95\% percentile (nonparametric) bootstrap-based confidence intervals.\vspace{0.5cm}}
\label{Fig1}
\end{figure}

The estimated relative $q$-susceptibility for the whole period is presented in Fig~\ref{Fig1} and implies that very young children in age group $[0,6)$ are about 0.182 (95\% percentile bootstrap-based CI: 0.146-0.230) times as susceptible compared to the first adult age class $[18,30)$ with relative susceptibility equal to 1 (95\% CI: 0.829-1.252). Primary school students aged $[6,12)$ have a relative susceptibility of 0.550 (95\% CI: 0.427-0.629) and secondary school students aged $[12,18)$ a susceptibility of 0.603 (95\% CI: 0.536-0.700). This shows an increasing $q$-susceptibility by increasing age up to $[18,30)$ after which the relative $q$-susceptibility tends to decrease slightly. Note, however, that this $q$-susceptibility captures not only differences in immunological susceptibility to infection and the rather low relative $q$-susceptibility in the $[12,18)$ age class could therefore be influenced by (compliance to) non-pharmaceutical interventions, over and above the age-specific contact frequencies. 

\begin{figure}[h!]
\centering
\includegraphics[width=1\textwidth]{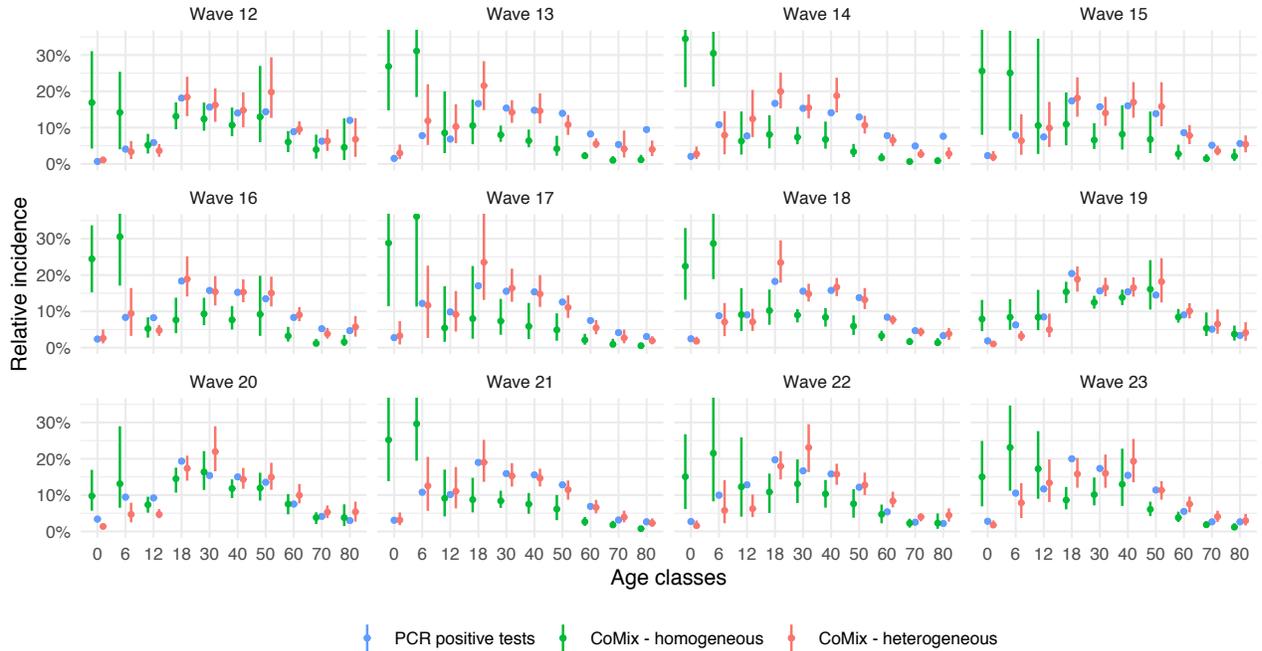}  
\caption{{\bf Estimated relative incidence.} In blue: relative incidence based on Belgian PCR data. In green: estimated relative incidence based on the next generation principle under the assumption of a homogeneous proportionality factor. In red: estimated relative incidence based on the next generation principle with an estimated age-specific $q$-susceptibility under the assumption on age-specific infectiousness as in Fig~\ref{Fig1}. Dots represent means and bars represent 95\% percentile (nonparametric) bootstrap-based confidence intervals.\vspace{0.5cm}}
\label{Fig2}
\end{figure}

The comparison of the relative incidence as estimated based on the positive PCR test data and the CoMix social contact data is presented in Fig~\ref{Fig2}. The social contact data are presented by waves starting with wave 12 on 22 December 2020 and an inter-survey wave interval of two weeks for subsequent waves (cf. details in Table~\labelcref{table_CoMix} of \nameref{S1_Appendix}). The nationally collected data are represented in blue and the estimates coming from social contact data in two colors: in green, the initial estimate with a homogeneous proportionality factor (i.e., with $a_i=1$ and $h_j=1$ for all $i,j$) and in red, the estimate using heterogeneous $q$-susceptibility and infectiousness as presented in Fig~\ref{Fig1}. We clearly observe that estimates of the relative incidence under the homogeneous proportionality factor assumption (green) are very different from the empirical estimates (blue), especially for the young age groups. The relative incidence among adult age classes is estimated relatively well up to a constant, but the relative incidence for children coming from the homogeneous social contacts approach is clearly overestimated, except perhaps during times of school closure (see, e.g., wave 19). This finding provides a clear indication that SARS-CoV-2 transmission is different in children as compared to adults and that homogeneity assumptions should be avoided given that such assumptions could lead to erroneous projections.\vspace{1cm}

\begin{figure}[h!]
\centering
\includegraphics[width=0.7\textwidth]{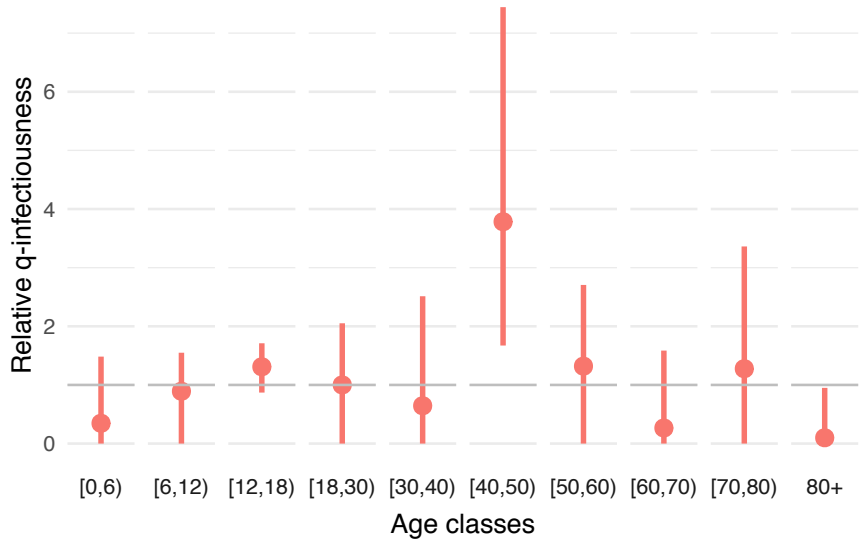}  
\caption{{\bf Estimated relative $q$-infectiousness.} The estimation of $q$-infectiousness is performed by age class using the next generation principle under an assumption on age-specific susceptibility $(0.4,0.39,0.38,0.79,0.86,0.8,0.82,0.88,0.74,0.74)$. The calibration is performed on CoMix waves 12 to 23. Dots represent means and bars represent 95\% percentile (nonparametric) bootstrap-based confidence intervals.\vspace{1cm}}
\label{Fig3}
\end{figure}

The result of estimating the $q$-infectiousness for the whole period, is depicted in Fig~\ref{Fig3}. The estimates also show a potential important heterogeneity concerning the proportionality factor on the infectiousness side. However, this reverse exercise provides less accurate results, with very large confidence intervals and some bootstrap estimates reaching zero, both being problematic when dealing with relative values. Those effects are the result of a lack of constraints for q-infectiousness. Indeed, while it is impossible to reach zero susceptibility for a specific age class when having at the same time non-zero incidence, it is technically allowed for age-specific infectiousness to be zero as the observed incidence could result from transmission from other age classes.

Exact estimates of the components of the $q$-susceptibility $(a_i)$ and $q$-infectiousness $(h_i)$ are provided in Tables \labelcref{sus1asstable,inf1asstable} of \nameref{S1_Appendix}. Additional estimates under the assumption of homogeneity regarding infectiousness or susceptibility (i.e., estimating $(a_i)$ under $h_j=1, \;\forall j$ or estimating $(h_j)$ under $a_i=1, \;\forall i$) are also presented in Figs~\labelcref{sus1homo,inf1homo} of \nameref{S1_Appendix} together with estimated values and the effect on the relative incidence. These additional estimates provide qualitatively similar results. Additional sensitivity analyses with regard to $(h_j)$ and $(a_j)$ showed that the variation of $q$-susceptibility estimates under different assumptions is clearly limited while sensitivity is greater concerning $q$-infectiousness estimates (see Figs~\labelcref{sensitivity_1,sensitivity_2} of~\nameref{S1_Appendix}). 

 \vspace{2cm}
\subsection*{Time evolution of proportionality factors}
 
Since proportionality factors capture several effects, they also capture time-dependent effects such as the reduction in susceptibility and infectiousness as a result of the vaccination campaign. In order to account for such a time evolution, we also performed the previous analysis using groups of two consecutive CoMix waves instead of the full period. The decision to consider two CoMix waves (28 days) together is motivated by the fact that a sufficiently long non-holidays period is required as social contacts in children are of importance and the heterogeneity of the transmission concerning adult classes is partially constrained by infection reported by children. Note that the gradual introduction of the alpha variant of concern might also interfere.
 
\begin{figure}[h!]
\centering
\includegraphics[width=0.95\textwidth]{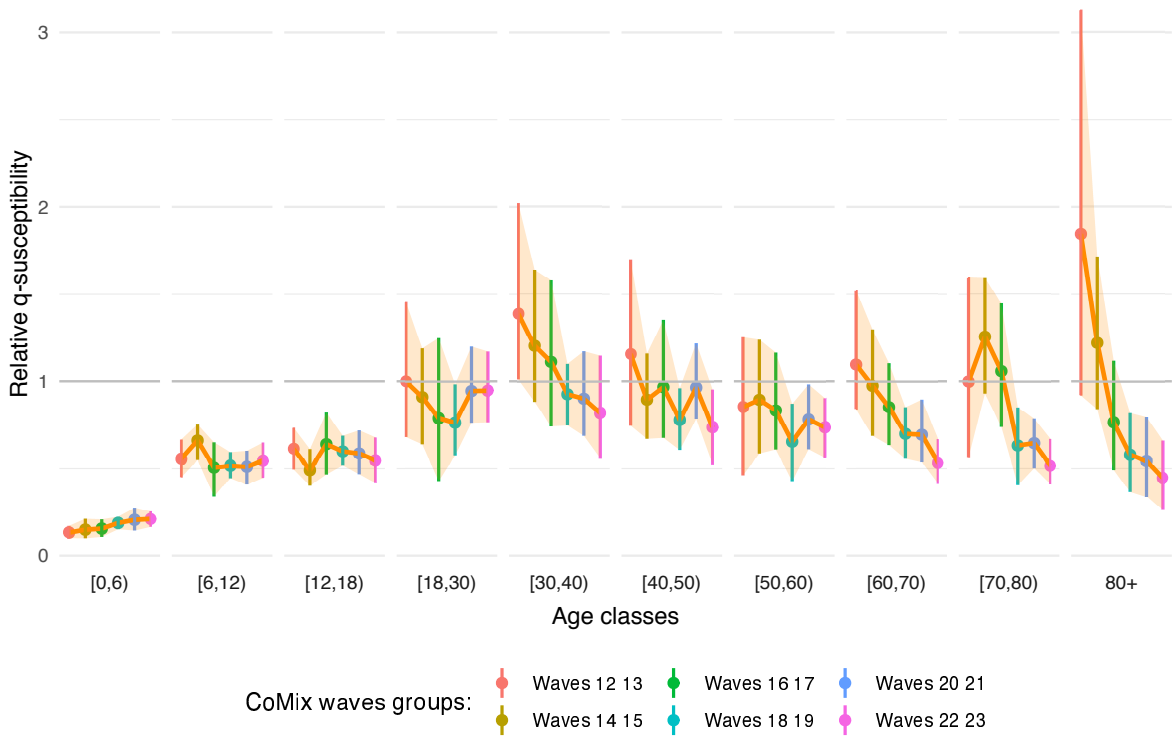}  
\caption{{\bf Estimated relative $q$-susceptibility over time.} The estimation of $q$-susceptibility is performed by age class and over time on groups of two consecutive CoMix waves, corresponding to a period of 4 weeks, under an assumption on age-specific infectiousness $(0.54,0.55,0.56,0.59,0.7,0.76,0.9,0.99,0.99,0.99)$. Dots represent means and bars represent 95\% percentile (nonparametric) bootstrap-based confidence intervals.\vspace{0.5cm}}
\label{Fig4}
\end{figure}

Estimates of the time-dependent $q$-susceptibility relying on the same (time-invariant) assumption with regard to the infectiousness vector are presented in Fig~\ref{Fig4}. A normalization of the relative values was performed over the different waves such that the average of the estimated factors for the age classes $[0,6)$, $[6,12)$ and $[12,18)$, i.e. $\frac{a_1+a_2+a_3}{3}$, is assumed constant. This choice is motivated by the fact that the vaccination campaign was not including children during the entire study period, hence proportionality factors regarding susceptibility can be expected to be more stable for these age classes (however still influenced by the evolution of the proportion of susceptible individuals due to the ongoing epidemic). Thus, the results provide an estimate of the evolution of adults' proportionality factors under an on average constant assumption for children $[0,18)$. A second normalization (global scaling) is performed under the assumption of a mean susceptibility of one for the first adult class $[18,30)$ for wave 12.

\newpage
As indicated previously concerning the estimation during the complete time-period, a decreasing $q$-susceptibility is observed through adult age classes (see Fig~\ref{Fig1}), with the oldest age class (80+) being the least susceptible among all adults aged 18 years or older. This is a priori in contrast with usual assumptions regarding age-specific susceptibility to SARS-CoV-2 infection. However, in Fig~\ref{Fig4}, we clearly observe the highest relative $q$-susceptibility in the $80+$ age class for the earliest waves as compared to all other age groups, or at least an equal  $q$-susceptibility by considering the lower side of the confidence interval.  Moreover, the $q$-susceptibility in the oldest age class decreases rapidly over time towards the lowest relative $q$-susceptibility equal to 0.446 (95\% CI: 0.266-0.660) among the adult age groups. In general, the estimated $q$-susceptibility is almost similar across the different adult classes during the first period with the exception of the oldest class $80+$ with an estimated relative $q$-susceptibility of 1.844 (95\% CI: 0.920-3.127). Overall, $q$-susceptibility estimates of other age classes tend to decrease over time, albeit at a slower pace and to a lesser extent. This is in line with the implementation of the vaccination policy in Belgium, giving vaccination priority to residents of nursing homes (CoMix waves 13-16, see schematic timeline in Fig~\labelcref{timeline} of \nameref{S1_Appendix}) and the elderly in the general population (CoMix waves 18-21), while going gradually down from old to young throughout the study period.

Exact values of the estimates in Fig~\ref{Fig4} are provided in Table~\labelcref{sus6asstable} of \nameref{S1_Appendix} as well as time-dependent $q$-susceptibility and $q$-infectiousness under the various constraints mentioned above.

\vspace{1cm}
\clearpage
\subsection*{Time evolution of $R_t$ using contact patterns}

In order to check the utility and validity of the use of a heterogeneous proportionality factor, we illustrate its application by determining the reproduction number $R_t$, or more specifically the variation of the reproduction number over time, and comparing this evolution with $R_t$ directly estimated from confirmed cases/hospitalizations data.

\begin{figure}[h!]
\centering
\includegraphics[width=1\textwidth]{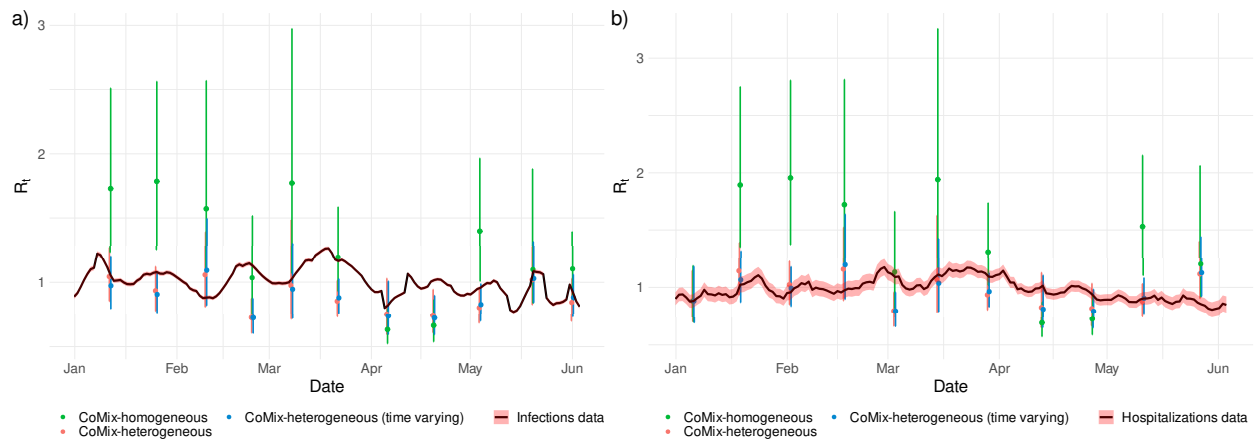} 
\caption{{\bf Temporal reproduction number.} Reproduction number estimated from the CoMix data using the next generation approach in comparison to the reproduction number estimated from the number of confirmed cases (a) and from the number of hospitalizations (b). In green: estimated $R_t$ under the assumption of a homogeneous proportionality factor. In red: estimated $R_t$ with the estimated age-specific $q$-susceptibility under the assumption on age-specific infectiousness as in Figs~\ref{Fig1} and \ref{Fig2}. In blue: estimated $R_t$ with the estimated temporal and age-specific $q$-susceptibility under the assumption on age-specific infectiousness as in Figs~\ref{Fig4}. Dots represent means and bars represent 95\% percentile (nonparametric) bootstrap-based confidence intervals.\vspace{0.5cm}}
\label{Fig5}
\end{figure}

In Fig~\ref{Fig5}, the variation of the reproduction number computed from the CoMix data is compared to the reproduction number computed either from the number of cases \cite{DSI} (panel a) or from hospitalizations \cite{Sciensano} (panel b). Clearly, specific choices of $q-$susceptibility and $q-$infectiousness affect the computation and in Fig~\ref{Fig5} we report results for the homogeneity scenario, heterogeneity scenario (corresponding to Fig~\ref{Fig1}) and temporal heterogeneity scenario (corresponding to Fig~\ref{Fig4}). A homogeneity assumption for $q$-infectiousness and $q$-susceptibility leads to a poor agreement with the reproduction number estimated from both confirmed cases and hospitalizations and is also characterized by a larger uncertainty. The use of the estimated heterogeneous reproduction factor agrees more with reality, with the use of temporal values is not leading to a substantial improvement.

\clearpage
\section*{Discussion}

We have demonstrated in this paper that social contact data can be used to inform transmission parameters and to estimate age-specific characteristics of SARS-CoV-2 transmission. More specifically, the next generation approach enables us to disentangle age-specific differences in transmission rates while relying on temporal changes in social contact behavior measured using consecutive waves of a social contact panel study. Clearly, SARS-CoV-2 transmission is partly influenced by age-specific differences in contact behavior, but importantly, additional age-specific factors related to susceptibility and infectiousness, in a broad sense, are necessary to account for. We have shown that such factors imply a smaller susceptibility for children as compared to adults, with the estimated susceptibility in children being around half of the susceptibility in adults, and even less for very young children (Fig~\ref{Fig1}). This result is in accordance with results obtained using CoMix social contact data in England but using a calibration on the reproduction number instead of the next generation approach \cite{Munday:2021wd} as well as in accordance with results obtained from more standard statistical methods \cite{Goldstein:2020ui,10.1001/jamapediatrics.2020.4573,Hu:2021ur,doi:10.1126/science.abe2424}. With respect to that, we assessed the impact of assuming homogeneous transmission parameters on the reproduction number, showing how (age-)heterogeneous parameters are necessary to correctly align the reproduction number from the CoMix data and the reproduction number estimated from infections or hospitalizations. Moreover, our method is able to estimate temporal transmission parameters and it shows a gradual decrease in susceptibility of adults in line with the progression of the Belgian vaccination campaign (Fig~\ref{Fig4}). This decrease implies a progressive change in the dynamics of the epidemic with largely unvaccinated childhood age groups gradually becoming more important drivers of SARS-CoV-2 transmission than predominantly vaccinated adult age groups.

However, our method suffers from several limitations. A potential bias which needs to be acknowledged is the use of PCR data which correspond to the observed relative incidence and do not necessarily correspond to the true relative incidence as each age class is not necessarily tested in the same way, even if we discard periods of strong variation in testing policy. Indeed, even in the absence of a change in testing policy (cf. Table~\labelcref{table_PCR} of \nameref{S1_Appendix}), age-specific differences in symptomatology, disease severity and the probability of developing symptoms upon infection lead to different shares of symptomatic and asymptomatic cases to be detected. Other approaches have also been investigated, for example using serological survey data instead of PCR data, but this was not successful on Belgian data given the limited amount of data and the poor synchronization between CoMix and serological survey periods. Moreover, using serological data requires addressing the difficulty of waning humoral immunity against SARS-CoV-2 infection. Despite the fact that we can infer $q$-susceptibility and $q$-infectiousness from the observed PCR test data, we cannot further disentangle both components by estimating the aforementioned quantities simultaneously. By comparing the two separate approaches, the estimation of the relative $q$-susceptibility seems most informative, since proportionality factors are better constrained by the data (cf.~also sensitivity analysis in Figs~\labelcref{sensitivity_1,sensitivity_2} of \nameref{S1_Appendix}). More specifically, the estimated $q$-susceptibility was identifiable when fitting to reported incidence data while $q$-infectiousness estimates were estimated to be zero for certain age classes, which seems an artifact of the methodology (which could potentially be solved by using external constraints on $q$-infectiousness to avoid reaching unrealistic low values of transmission). Another limitation of the proposed method is that a further decomposition of $q$-susceptibility (or $q$-infectiousness) in immunological susceptibility (infectiousness) and other external factors relevant for transmission between susceptible and infectious persons is difficult, at least without availability of relevant additional data thereon. Nonetheless, an assessment and quantification of the (relative) $q$-susceptibility, $q$-infectiousness and the corresponding relative incidence provides useful insights into heterogeneous SARS-CoV-2 transmission dynamics.  
\vspace{1cm}

\clearpage
\section*{Materials and methods}\label{MandM}
\subsection*{Social contact data}

Our study is based on Belgian social contact data collected within the CoMix survey \cite{Coletti2020,Verelst2021} during the COVID-pandemic between December 2020 and May 2021. These data are stored, processed and stratified by age by means of the online Socrates tool \cite{Verelst2021,Willem:2020uv,Socrates}. Participants were asked to fill in a contact dairy including all contacts made during a specific day, reporting the type of contact, location, and age of the contacted person, with a contact defined as an in-person conversation of at least a few words, or a skin-to-skin contact. The CoMix survey was repeatedly performed in different waves and different survey periods. More specifically, an initial survey period containing 8 waves was carried out between 4 March 2020 and 27 July 2020 targeting adults only. A second survey period, still ongoing in 2021, began on 11 November 2020 targeting participants of all ages. The waves are conducted with an interval of two weeks (14 days). For more detailed information on the CoMix survey and the stratification process, the reader is referred to \cite{Coletti2020,Verelst2021}. A detailed timetable of the CoMix waves and survey periods is presented in Table~\labelcref{table_CoMix} of \nameref{S1_Appendix}. A schematic timeline of CoMix waves according to the evolution of the alpha variant of concern and vaccination campaign in Belgium is presented in Fig~\labelcref{timeline} of \nameref{S1_Appendix}.

We use the following notation. $N_{i}$ denotes the number of individuals in the Belgian population of age $i$ according to Belgian demographic data \cite{StatBel} and integrated into the Socrates tool \cite{Socrates}. In general, we use subscripts $i$ as an index for the participant's age, and $j$ as an index for the contacted person's age. The following observable quantities (dependent on the wave chosen) can be extracted from the survey:
\begin{itemize}
\item $m_{ij}$ represents the average daily number of individuals of age $j$ who are contacted by a participant of age $i$. The elements $m_{ij}$ constitute a matrix $\mathbf{M}$ called \emph{social contact matrix}.
\item $c_{ij}$ is the per capita contact rate per day for participants of age $i$ with persons of age $j$ in the population. The elements $c_{ij}$ constitute a matrix $\mathbf{C}$ called the \emph{contact rate matrix}. This matrix is related to the social contact matrix by the relation $c_{ij} = m_{ij}/N_j $.
\end{itemize}

In theory, due to the reciprocal nature of contacts, the total number of contacts between members of two age classes, as reported by participants in each of the age groups, must be equal, hence $N_i m_{ij} = N_i c_{ij} N_j = N_j c_{ji} N_i = N_j m_{ji}$, which is equivalent to the condition that the contact rate matrix should be symmetric, i.e., $c_{ij} = c_{ji}$, $\forall i,j$. The social contact matrix $\mathbf{M}$ respects the relation $N_i m_{ij} = N_j m_{ji}$, but is in general not symmetric due to differences in $N_i$ and $N_j$. In practice, the observed total number of contacts $N_i m^{\text{raw}}_{ij}$ and $N_j m^{\text{raw}}_{ji}$ are not necessarily equal due to sampling bias, hence, we calculate the reciprocal social contact matrix by:
\[ m_{ij} = \frac {m^{\text{raw}}_{ij} N_i + m^{\text{raw}}_{ji} N_j}{2 N_i}\cdot\]

All these notations and definitions are similar to those described in detail in \cite{Handbook}, except that the subscripts $i$ and $j$ and order of indices are inverted here such that the definition of the social contact matrix $\mathbf{M}$ corresponds to the default output of the Socrates tool \cite{Socrates}.

\subsection*{Next generation principle}
The \emph{social contact hypothesis} \cite{Wallinga2006} implies that the age-specific number of infectious contacts is proportional to the self-reported age-specific number of social contacts. There are two ways to interpret empirical social contact survey data in light of this hypothesis: either survey participants can be infected by their infectious contacts or participants can infect their susceptible contacts. Here, we consider the first interpretation as initial definition -- since the CoMix survey did not specifically target infected persons, and symptomatic participants may have been less likely to participate in the survey. However, we will show that the two interpretations lead to the same mathematical result under the assumption of reciprocity of social contacts.

If we denote by $w_j$ the incidence within age class $j$ over a short observation interval (e.g. corresponding to a wave period), then $v_j = w_j / N_j$ is the risk of being infected during the observation interval for that age class (incidence rate or force of infection). The new generation of infected people is given by:
\[ w^\prime_i = \sum_j q \,N_i\, m_{ij} \,v_j = \sum_j  q \,m_{ij} \frac{N_i }{N_j } \,w_j, \]
where $q$ is a general proportionality factor completely defining the relationship between infection and contact events. The $q$-factor accommodates several effects such as susceptibility to infection, infectiousness upon infection, duration of the infectious period, type and effectiveness of contacts, seasonality, pre-existing natural and vaccine-induced immunity, etc.

The elements $k_{ij} =  q \,m_{ij} \frac{N_i }{N_j} = q N_i c_{ij}$ define a matrix $\mathbf{K}$ called the \emph{next generation matrix} (or \emph{reproduction matrix}) since $k_{ij}$ represents the mean number of individuals of age $i$ that are infected through a single individual of age $j$ during their entire infectious period (for which the time between consecutive generations of infected individuals is chosen to be equal to the average duration of infectiousness). 

Note that under the reciprocity assumption leading to a symmetric matrix $\mathbf{C}$, the relation $N_i m_{ij} = N_j m_{ji}$ provides:
\[ k_{ij} =  q \,m_{ij} \frac{N_i }{N_j } = q \, m_{ji} \qquad\text{or}\qquad \mathbf{K} = q \,\mathbf{M}^T,\]
corresponding to the second interpretation that survey participants (on the right side of the transpose contact matrix $\mathbf{M}^T$) can directly infect their contacts (now on the left side) modulo the proportionality factor. This expression relying on the transpose of the social contact matrix obtained as a direct output of the Socrates tool, $\mathbf{M}$, is chosen because of its better numerical stability.

The recurrence relation of the next generation matrix $\mathbf{K}$:
\[ w^{\prime\prime} = \mathbf{K} \,w^\prime =   \mathbf{K}^2\, w, \qquad w^{\prime\prime\prime} =  \mathbf{K}^3\, w, \qquad \dots\]
tends to a stable distribution due to the Perron--Frobenius theorem \cite{meyer2000matrix}, i.e.,
\[R_{t}\, w^* = \mathbf{K} \,w^* = q \,\mathbf{M}^T \,w^*,\] with $R_t$ corresponding to the reproduction number of SARS-CoV-2 \cite{Diekmann:1990tk} which is defined as the leading eigenvalue of the next generation matrix $\mathbf{K}$. More specifically, estimation of the reproduction number $R_{t}$ and the relative incidence $w$ can be done by computing the leading eigenvalue and corresponding right-eigenvector of $\mathbf{K}$. However, $R_{t}$ depends on the proportionality factor $q$, which might be unknown, but the relative incidence $w$ is independent of $q$ and can therefore be directly extracted from the social contact data $\mathbf{M}^T$. The reproduction number is initially the basic reproduction number $R_0$, but switches to the effective reproduction number as long as social contact data evolve and the proportionality factor $q$ captures the depletion of susceptible. We emphasize here that the eigenvector $w$ is only recovered up to a global constant and therefore individual components $w_j$ have no meaning. What can be interpreted are relative ratios such as  $w_i/w_j$, providing an estimate of the relative incidence in age class $i$ as compared to the incidence in age class $j$. This vector is usually normalized such that $\sum_i w_i = 1$. In the same way, the incidence rate $v_i$ can be recovered, in relative sense, as the leading left-eigenvector of $\mathbf{M}^T$.

The switch from a homogeneous proportionality factor $q$ to a heterogeneous $q_{ij}$ is performed by assuming:
\[ q_{ij} = \tilde q \, a_i \, h_j \]
where the vector $(a_i)$ acts on the susceptible side,  the vector $(h_j)$ acts on the infectiousness side, and $\tilde q$ is a remaining global proportionality factor  captures any residual effect. This remaining factor has no influence on the computation of the relative incidence $w$. However, due to the presence of $\tilde q$, the vectors $(a_i)$ and $(h_j)$ only have a relative interpretation.

The heterogeneous next generation matrix is defined as:
\[ k_{ij} =  \tilde q \,a_i\,m_{ij} \frac{N_i }{N_j }\,h_j  \qquad\text{or}\qquad \mathbf{K} = \tilde q\,\diag{a_i} \,\mathbf{M}^T\diag{h_j}.\]

We note that we are working here with a next generation matrix with \emph{small domain}. There also exists a next generation matrix with \emph{large domain} taking explicitly into account the different states of the disease and their duration for each age class \cite{diekmann2000mathematical}. However, the small domain approach is appropriate here since we do not work with a dynamical system and heterogeneity in disease duration is part of the effects captured by the proportionality factors.

\subsection*{Estimating relative $q$-susceptibility $(a_i)$ and $q$-infectiousness $(h_j)$ from COVID-19 age-structured indicators}

The vectors $(a_i)$ and $(h_j)$ have an important impact on the determination of the leading right eigenvector in the system:
\[  \mathbf{K} \, w^*=  \tilde q\,\diag{a_i} \,\mathbf{M}^T\diag{h_j} \, w^* = R_t \,w^*. \]
The obtained relative incidence $w^*$ can be compared with the normalized relative incidence $ \tilde w$ estimated from the observed incidence data in Belgium. Using this approach, we are able to determine $q$-susceptibility and $q$-infectiousness corresponding to SARS-CoV-2 transmission in Belgium. However, $(a_i)$ and $(h_j)$ vectors cannot be estimated simultaneously in a unique way from this process since there remains an indeterminacy \cite{FLASCHE2011125,Goeyvaerts}. Nevertheless, the identifiability problem can be solved by imposing a constraint on one of the two vectors.

For this study, we choose each time a heterogeneous constraint coming from the literature as well as a homogeneous constraint (whose results are only presented in \nameref{S1_Appendix}). The heterogeneous constraints are defined from the following assumptions:
\begin{itemize}
\item For the assumption on infectiousness $h_j$ (estimation of $q$-susceptibility parameters $a_i$):  We consider the probability of an asymptomatic COVID-19 infection in case of SARS-CoV-2 exposure in the Belgian population to be $$p = (0.94, 0.92, 0.90, 0.84, 0.61, 0.49, 0.21, 0.02, 0.02, 0.02)$$ as assumed in \cite{ABRAMS2021100449} using data from \cite{Wu:2020tk}. Assuming that the relative infectiousness of asymptomatic versus symptomatic individuals is $0.51$ \cite{ABRAMS2021100449}, we obtain the following constraint:
$$(h_j) = 0.51 p + 1 (1-p) = (0.54,0.55,0.56,0.59,0.7,0.76,0.9,0.99,0.99,0.99).$$
\item For the assumption on susceptibility $a_i$ (estimation of $q$-infectiousness parameters $h_j$): The assumption is taken from \cite{Davies:2020wt}:
$$(a_i) = (0.4,0.39,0.38,0.79,0.86,0.8,0.82,0.88,0.74,0.74).$$
\end{itemize}

\subsection*{Data and fitting procedure}

We use Belgian data on daily incidence of COVID-19 confirmed by means of a positive PCR test, as provided by the Belgian Institute for Public Health, Sciensano \cite{Sciensano}. In order to reduce testing biases, the period of study is restricted to a period with almost constant testing policy (mandatory testing for both symptomatic cases and asymptomatic close contacts or red zone travelers) and before biases are induced by the introduction of the EU Digital COVID Certificate, see \cite{Sciensano_PCR} or Table~\labelcref{table_CoMix} of \nameref{S1_Appendix}) for a summary). Since there is a delay between a change in social contact behavior and its effect on the relative incidence, we consider PCR test results for the period starting 7 days after the onset of a specific CoMix wave and lasting for 14 days thereafter. 

Concerning social contact data, the initial CoMix survey waves (1 to 8) are discarded due to a variable testing policy and lack of information regarding child-child contacts. The three subsequent waves (9 to 11) are also discarded since, despite the introduction of measuring child-child contacts, the information was collected using a different survey formulation. CoMix waves 12 to 23 correspond to a period with constant testing policy, an identical survey design as well as without vaccination in children, which implies that the results with regard to age classes $[0-6)$, $[6-12)$ and $[12-18)$ years are expected to be more stable. The start of wave 12 corresponds to 22 December 2020 when the vaccination campaign in adults has not been started and the last wave considered corresponds to 26 May 2021, with PCR tests considered up to 15 June 2021 (thus when vaccination of the oldest individuals in the Belgian population was nearly completed).

The estimation of the parameters $a_i$ or $h_i$ is performed using the statistical software package~R. A minimum-distance estimation is performed using the Hellinger distance~\cite{Hellinger} (which is suitable for distributions) between relative incidences $w^*$ and $\tilde w$. The optimization is done by means of a random search numerical method \cite{rastrigin1963convergence}, starting from an initial homogeneous prior ($a_i=1$ or $h_i=1$, $\;\forall i=1\dots10$). The process uses a Gaussian random walk with steps of length $\mathcal{N}{\left(0,0.005^2\right)}^{10}$ which are performed until no change in the distance is observed during $100$ consecutive iterations. The sensitivity analysis is provided by repeating the process over $200$ nonparametric bootstrap runs using the previous posterior as new prior. Uncertainty is quantified using means and 95\% percentile confidence intervals (i.e., $2.5\%$ and $97.5\%$ quantiles of all bootstrap-based estimates). 

Since $q$-susceptibility and $q$-infectiousness vectors represent relative values, a normalization process should be chosen for the representation of the results. We choose here the following (two-step) normalization process:
\begin{itemize}
\item The average $\frac{a_1+a_2+a_3}{3}$ or $\frac{h_1+h_2+h_3}{3}$ of the estimated factors for the age classes [0,6), [6,12) and [12,18) (children, no subject to vaccination during the complete period) is assumed to be constant across all bootstrap runs and wave groups if applicable (i.e. all combinations of two CoMix waves).
\item The mean $\overline{a_4}$ or $\overline{h_4}$ (across the bootstrap runs) for the first adult age class $[18,30)$ is set to one. This constraint is chosen because we mainly want to compare susceptibility and infectiousness of children versus adults while the age class $[18,30)$ is one of the most stable ones in the bootstrapping process. Note that this second normalization step is only a global scaling, thus conserving the confidence interval around one for the age class $[18,30)$.
\end{itemize}

\subsection*{$R_t$ evolution from contact patterns}
Via the next generation approach, the ratio of the eigenvalues of two next generation matrices can be used to evaluate the relative reduction in the reproduction number. This can be done to compare the temporal reproduction number derived from the CoMix survey with independent evaluations of the reproduction number. We use as comparison the $R_t$ computed from the daily number of cases \cite{DSI} and the daily number of hospitalizations \cite{Sciensano}. In order to account for the time delays associated with infections and hospitalizations (e.g. time to develop symptoms, time to hospitalizations, etc.), the reproduction number computed from the CoMix social contact data was shifted forward in time. A time shift of 7 (14) days is considered when comparing $R_t$ estimates with the reproduction number computed from the number of confirmed cases (respectively hospitalizations). These time shifts are chosen in order to take account of a mean delay between infection and (symptomatic) testing as well as an additional delay between symptom onset and hospitalization~\cite{Faes2020}. As the reproduction number is known up to the overall constant $\tilde q$, we fix the reproduction number for CoMix wave 12 to be equal to the reproduction number computed from infections or hospitalizations. Uncertainty due to sampling variability is estimated via 10000 nonparametric bootstraps.

\clearpage
\section*{Supporting information}

\paragraph*{S1 Appendix.}
\label{S1_Appendix}
{\bf Supplementary Material.} Presentation of the complete results with all figures and tables containing values. Table A: Timetable of CoMix starting dates. Fig A. Schematic timeline of CoMix waves. Table B: Timetable of Belgian testing policy. Figs. B and C. Sensitivity analysis of the method under different assumptions.  Figs. D to S: Estimates of relative proportionality factors and corresponding relative incidence under different assumptions. Table C to J: Exact estimates of the components of the proportionality factors.

\section*{Acknowledgments}
We thank several researchers from the SIMID COVID-19 consortium (interuniversity collaboration between University of Antwerp (CHERMID) and UHasselt (DSI, CenStat) as well as other researchers from the Interuniversity Institute of Biostatistics and statistical Bioinformatics (I-BioStat) (KU Leuven and UHasselt) for numerous constructive discussions and meetings. The authors thank the EpiPose consortium partners for useful discussions and for help in setting up the CoMix survey as part of EpiPose. The authors are also very grateful for access to the data from the Belgian Scientific Institute for Public Health, Sciensano.

\section*{Author contributions}  
Conceptualization: NF, SA, FB, NH; Literature research: NF, LA, AL; Data preparation and/or collection: PC, LW; Data analysis code: NF, PC; Coordination: NH; All authors interpreted the findings, contributed to writing the manuscript, the revised manuscript and approved the final version of the manuscript for publication.

\section*{Financial support} 

This work received funding from the European Research Council (ERC) under the European Union’s Horizon 2020 research and innovation program -- Project TransMID (NF, PC and NH, Grant Agreement 682540). This project was also funded by the European Union’s Horizon 2020 Research and Innovations Programme -- Project EpiPose, Epidemic intelligence to Minimize COVID-19’s Public Health, Societal and Economical Impact (NH, grant number 101003688). LW gratefully acknowledges support from the Research Foundation Flanders (FWO) (postdoctoral fellowships 1234620N). NH acknowledges funding from the Antwerp Study Center for Infectious Diseases (ASCID) and the Methusalem-Centre of Excellence consortium VAX–IDEA.  The funders had no role in the study design, data collection and analysis, decision to publish or preparation of the manuscript.

\section*{Competing interests}

None.

\section*{Data availability statement} 

R codes and all necessary data to run the codes (potentially aggregate) are available on GitHub at \url{https://github.com/nicolas-franco-unamur/Next-gen}. CoMix data and age-class re-aggregate PCR tests data are provided with the code. CoMix social contact data are also available via \url{https://www.socialcontactdata.org}. If needed, initial non-aggregate PCR tests data can be requested from Sciensano via the online form \url{https://epistat.wiv-isp.be/datarequest}.

\nolinenumbers

%
%
%
\bibliography{references}

%
%
%
%

\label{maindocumentlastpage}
\clearpage
\rfoot{\thepage/\pageref{LastPage}}
\renewcommand{\footrule}{}
\lfoot{}

\pagenumbering{arabic}
\newgeometry{top = 2.5cm, bottom = 4cm, left = 2.5cm, right =2.5cm}
\setlength{\oddsidemargin}{0in}
\setlength{\evensidemargin}{0in}

\renewcommand*{\thepage}{S\arabic{page}}
\appendix
\nolinenumbers
\linespread{1} 
\newcommand{\beginsupplement}{%
        \setcounter{table}{0}
        \renewcommand{\thetable}{\Alph{table}}%
        \setcounter{figure}{0}
        \renewcommand{\thefigure}{\Alph{figure}}%
     }
\beginsupplement
\singlespacing
\def\justifying{%
  \rightskip=0pt
  \spaceskip=0pt
  \xspaceskip=0pt
  \relax
}
\justifying
\captionsetup{font={small,stretch=1},justification=centering}

\section*{S1 Appendix: Supplementary Material}

In this Supplementary Material, we present additional results on the estimation of heterogeneous proportionality factors using the next generation matrix as well as some timeline descriptions and a sensitivity analysis to assumptions.\\

Social contact data $\mathbf{M}$ are taken from CoMix waves 12 to 23 as described in the timetable given in Table \labelcref{table_CoMix} and timeline in Fig~\labelcref{timeline} and PCR tests were performed under the testing policy presented in Table \labelcref{table_PCR}. Relative incidences are computed as the leading eigenvector of the next generation matrix $ \diag{a_i} \,\mathbf{M}^T\diag{h_j}  $ and compared with the normalized relative incidence estimated from PCR positive tests. The $q$-susceptibility $(a_i)$ and $q$-infectiousness $(h_j)$ vectors are either assumed or estimated. A sensitivity analysis concerning the assumptions on $(a_i)$ and $(h_j)$ vectors is presented in Figs~\labelcref{sensitivity_1,sensitivity_2}.\\

In the last subsections, we present the complete details of 4 different sets of results:
\begin{itemize}
\item Estimation of $q$-susceptibility $(a_i)$ using homogeneous infectiousness assumption $\; h_j=1 \;\forall j$ (Figs~\labelcref{sus1homo,sus1homong,sus6homo,sus6homong} and Tables \labelcref{sus1homotable,sus6homotable})
\item Estimation of $q$-susceptibility $(a_i)$ using heterogeneous infectiousness assumption $(h_j)$ coming from the literature (Figs~\labelcref{sus1ass,sus1assng,sus6ass,sus6assng} and Tables \labelcref{sus1asstable,sus6asstable})
\item Estimation of $q$-infectiousness $(h_j)$ using homogeneous susceptibility assumption $\; a_i=1 \;\forall i$ (Figs~\labelcref{inf1homo,inf1homong,inf6homo,inf6homong} and Tables \labelcref{inf1homotable,inf6homotable})
\item Estimation of $q$-infectiousness $(h_j)$ using heterogeneous susceptibility assumption $(a_i)$ coming from the literature (Figs~\labelcref{inf1ass,inf1assng,inf6ass,inf6assng} and Tables \labelcref{inf1asstable,inf6asstable})
\end{itemize}

Each time, we present the estimate of the proportionality factors for the complete period (waves 12 to 23) or estimated by groups of two waves, with the exact values presented in the subsequent tables. The normalization method is described at the beginning of each subsection. We associate to the results the estimate of the relative incidence, with the real data presented in blue, the estimate without any proportionality factor $a_i=h_i=1 \;\forall i$ in green and the estimate with the assumed and estimated proportionality factors in red. Dots represent means and bars represent 95\% nonparametric bootstrap confidence intervals ($2.5\%$ and $97.5\%$ quantiles). 

\clearpage

\subsection*{CoMix and PCR test timelines}

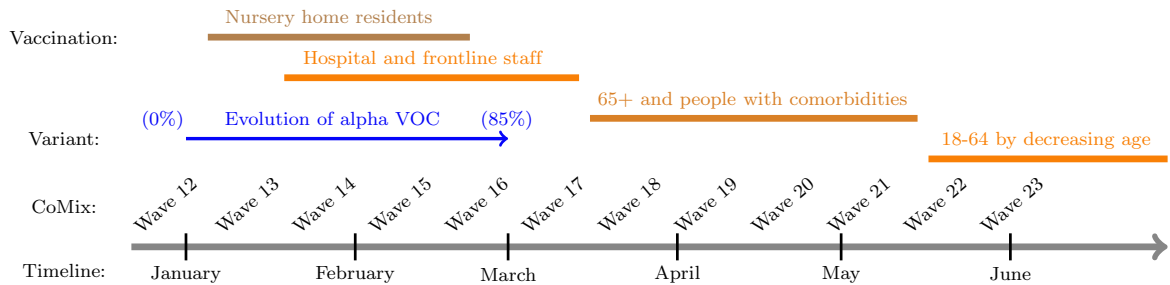
\begin{figure}[h!]
\begin{center}
\scalebox{0.9}{\footnotesize
\begin{tikzpicture}
\draw[line width=1mm,color=gray!95,->] (-10/31*2.5,0) -- (5.8*2.5,0) ;
\draw[line width=0.4mm,color=black] (0*2.5,0.2) -- (0*2.5,-0.2) ;
\draw[line width=0.4mm,color=black] (1*2.5,0.2) -- (1*2.5,-0.2) ;
\draw[line width=0.4mm,color=black] (1*2.5+28/31*2.5,0.2) -- (1*2.5+28/31*2.5,-0.2) ;
\draw[line width=0.4mm,color=black] (2*2.5+28/31*2.5,0.2) -- (2*2.5+28/31*2.5,-0.2) ;
\draw[line width=0.4mm,color=black] (2*2.5+58/31*2.5,0.2) -- (2*2.5+58/31*2.5,-0.2) ;
\draw[line width=0.4mm,color=black] (3*2.5+58/31*2.5,0.2) -- (3*2.5+58/31*2.5,-0.2) ;
\node at (0*2.5,-0.4){January};
\node at (1*2.5,-0.4){February};
\node at (1*2.5+28/31*2.5,-0.4){March};
\node at (2*2.5+28/31*2.5,-0.4){April};
\node at (2*2.5+58/31*2.5,-0.4){May};
\node at (3*2.5+58/31*2.5,-0.4){June};
\node at (-1.8,-0.35){Timeline:};
\node at (-1.8,0.6){CoMix:};
\node at (-1.8,3.1){Vaccination:};
\node at (-1.8,1.6){Variant:};
\node[right,rotate=40] at (-10/31*2.5,0.2){Wave 12};
\node[right,rotate=40] at (5/31*2.5,0.2){Wave 13};
\node[right,rotate=40] at (19/31*2.5,0.2){Wave 14};
\node[right,rotate=40] at (1*2.5+2/31*2.5,0.2){Wave 15};
\node[right,rotate=40] at (1*2.5+16/31*2.5,0.2){Wave 16};
\node[right,rotate=40] at (1*2.5+28/31*2.5+2/31*2.5,0.2){Wave 17};
\node[right,rotate=40] at (1*2.5+28/31*2.5+16/31*2.5,0.2){Wave 18};
\node[right,rotate=40] at (1*2.5+28/31*2.5+30/31*2.5,0.2){Wave 19};
\node[right,rotate=40] at (2*2.5+28/31*2.5+13/31*2.5,0.2){Wave 20};
\node[right,rotate=40] at (2*2.5+28/31*2.5+27/31*2.5,0.2){Wave 21};
\node[right,rotate=40] at (2*2.5+58/31*2.5+11/31*2.5,0.2){Wave 22};
\node[right,rotate=40] at (2*2.5+58/31*2.5+25/31*2.5,0.2){Wave 23};
\draw[line width=1mm,color=orange!40!gray] (4/31*2.5,3.1) -- (1*2.5+21/31*2.5,3.1)  node[above left] {Nursery home residents\ \ \ \ \ };
\draw[line width=1mm,color=orange!95!gray] (18/31*2.5,2.5) -- (1*2.5+28/31*2.5+13/31*2.5,2.5)  node[above left] {Hospital and frontline staff\ \ \ \ \ };
\draw[line width=1mm,color=orange!70!gray] (1*2.5+28/31*2.5+15/31*2.5,1.9) -- (2*2.5+58/31*2.5+14/31*2.5,1.9)  node[above left] {65+ and people with comorbidities\ };
\draw[line width=1mm,color=orange!95!gray] (2*2.5+58/31*2.5+16/31*2.5,1.3) -- (5.8*2.5,1.3)  node[above left] {18-64 by decreasing age\ \ };
\draw[line width=0.5mm,color=blue!95!gray,->] (0,1.6) -- (1*2.5+28/31*2.5,1.6)  node[above left] {(0\%)\qquad Evolution of alpha VOC\qquad(85\%)\!\!\!\!\!\!\!\!\!\!};
\end{tikzpicture}}
\end{center}
\caption{Schematic timeline of CoMix waves, vaccination campaign in Belgium and evolution of alpha VOC.}
\label{timeline}
\end{figure}

We present here details concerning Belgian timelines on social contact data collection and PCR testing policy. A detailed timetable of the CoMix waves and survey periods is presented in Table~\labelcref{table_CoMix}. The period of study is restricted on CoMix waves 12 to 23 since it corresponds to an identical survey design. A schematic timeline of CoMix waves according to the evolution of the alpha variant of concern and vaccination campaign in Belgium is presented in Fig~\labelcref{timeline}. This period of study also corresponds to an almost constant testing policy, with mandatory testing both symptomatic cases and asymptomatic close contacts and travelers and before biases induced by the introduction of the EU Digital COVID Certificate. A summary of Belgian testing policy is presented in Table~\labelcref{table_PCR} though more information can be found in \cite{Sciensano_PCR}.

\begin{table}[ht]
\centering\scriptsize
\begin{tabular}{|rl|rl|rl|}
  \hline
  \multicolumn{2}{|c|}{Survey 1} & \multicolumn{2}{c|}{Survey 2a} & \multicolumn{2}{c|}{Survey 2b}\\
  \hline
wave 1 & 20 April 2020 & wave 9 & 11 November 2020 & wave 12 & 22 December 2020\\
wave 2 & 5 May 2020 & wave 10 & 25 November 2020 & wave 13 & 6 January 2021\\
wave 3 & 18 May 2020 & wave 11 & 9 December 2020 & wave 14 & 20 January 2021\\
wave 4 & 1 June 2020 &  &  & wave 15 & 3 February 2021\\
wave 5 & 15 June 2020 &  &  & wave 16 & 17 February 2021\\
wave 6 & 29 June 2020 &  &  & wave 17 & 3 March 2021\\
wave 7 & 13 July 2020 &  &  & wave 18 & 17 March 2021\\
wave 8 & 27 July 2020 &  &  & wave 19 & 31 March 2021\\
 & &  &  & wave 20 & 14 April 2021\\
 & &  &  & wave 21 & 28 April 2021\\
 & &  &  & wave 22 & 12 May 2021\\
 & &  &  & wave 23 & 26 May 2021\\
   \hline
\end{tabular}
\caption{Timetable of CoMix with starting dates. The first survey (waves 1-8) did not include children. The subsequent waves included children with a preliminary version of the survey (waves 9-11), which is modified and finalized in December 2020 (waves 12-$\dots$).} 
\label{table_CoMix}
\end{table}

\begin{table}[ht]
\centering\scriptsize
\begin{tabular}{|c|l|}
  \hline
Date & Change in testing policy\\
  \hline
05 April 2020 & Start of mandatory testing of all symptomatic cases with PCR tests\\
06 June 2020 & Start of mandatory testing of all asymptomatic close contacts with PCR tests\\
21 October 2020 & Temporary stop of testing of all asymptomatic close contacts\\
22 November 2020 & Restart of mandatory testing of all asymptomatic close contacts\\
31 December 2020 & Mandatory testing of all red zone travelers (2 PCR tests: days 1 and 7) \\
25 January 2021 & Asymptomatic close contacts must perform 2 PCR tests (days 1 and 7) \\
16 June 2021 & Antigen and PCR tests are available for EU Digital COVID Certificates\\
   \hline
\end{tabular}
\caption{Summary of the main changes in Belgian testing policy.} 
\label{table_PCR}
\end{table}

\clearpage

\subsection*{Sensitivity analysis}

\begin{figure}[htb!]
\begin{center}
\includegraphics[width=0.6\textwidth]{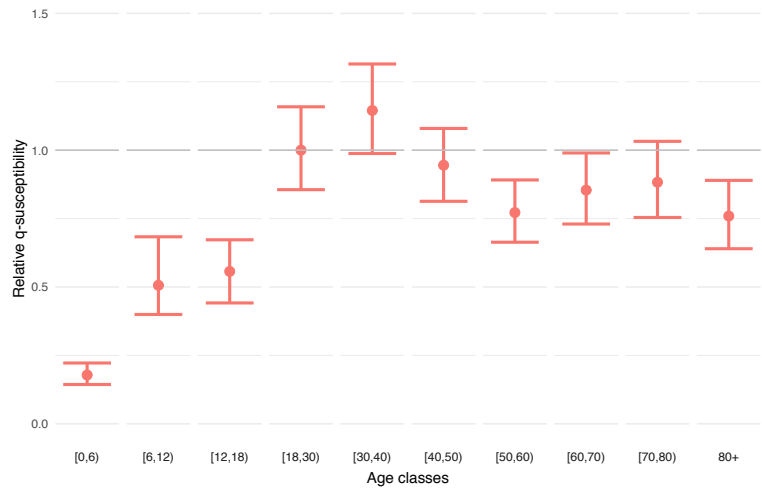}
\end{center}
\caption{Sensitivity analysis on the estimation of relative $q$-susceptibility under random heterogeneous uniform variation $\mathcal U^{10}_{[-0.1,0.1]}$ on the assumption on infectiousness {\scriptsize$(0.54,0.55,0.56,0.59,0.7,0.76,0.9,0.99,0.99,0.99)$}.}
\label{sensitivity_1}
\end{figure}

\begin{figure}[htb!]
\begin{center}
\includegraphics[width=0.6\textwidth]{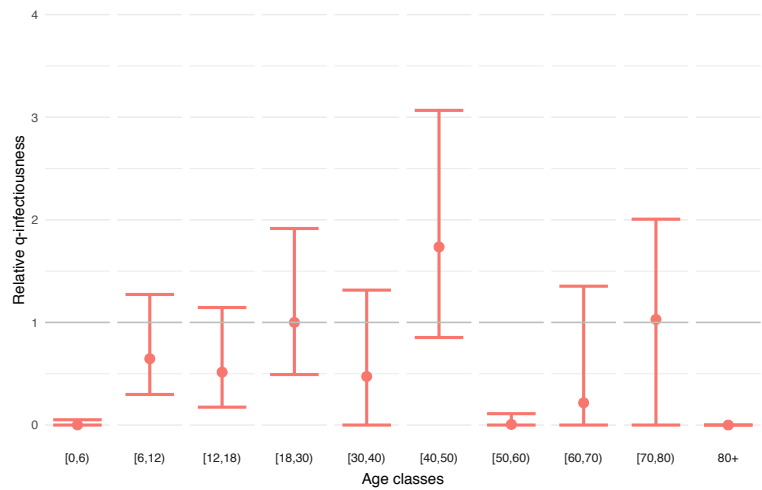}
\end{center}
\caption{Sensitivity analysis on the estimation of relative $q$-infectiousness under random heterogeneous uniform variation $\mathcal U^{10}_{[-0.1,0.1]}$ on the assumption on susceptibility {\scriptsize$(0.4,0.39,0.38,0.79,0.86,0.8,0.82,0.88,0.74,0.74)$}.}
\label{sensitivity_2}
\end{figure}

\begin{figure}[htb!]
\begin{center}
\includegraphics[width=0.5\textwidth]{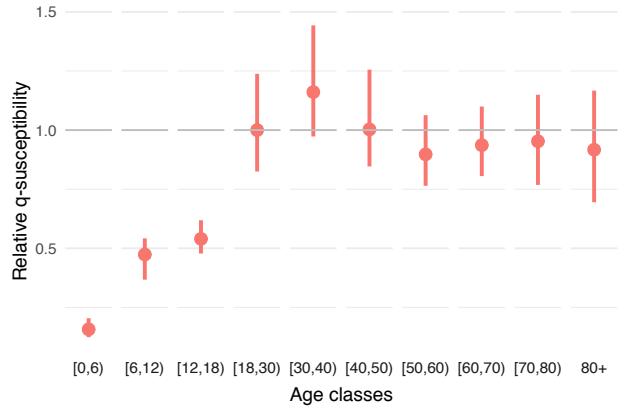}
\end{center}
\caption{Estimation of relative $q$-susceptibility using assumption on infectiousness {\scriptsize$(1,1,1,1,1,1,1,1,1,1)$}.}
\label{sus1homo}
\end{figure}

\begin{figure}[htb!]
\begin{center}
\includegraphics[width=0.9\textwidth]{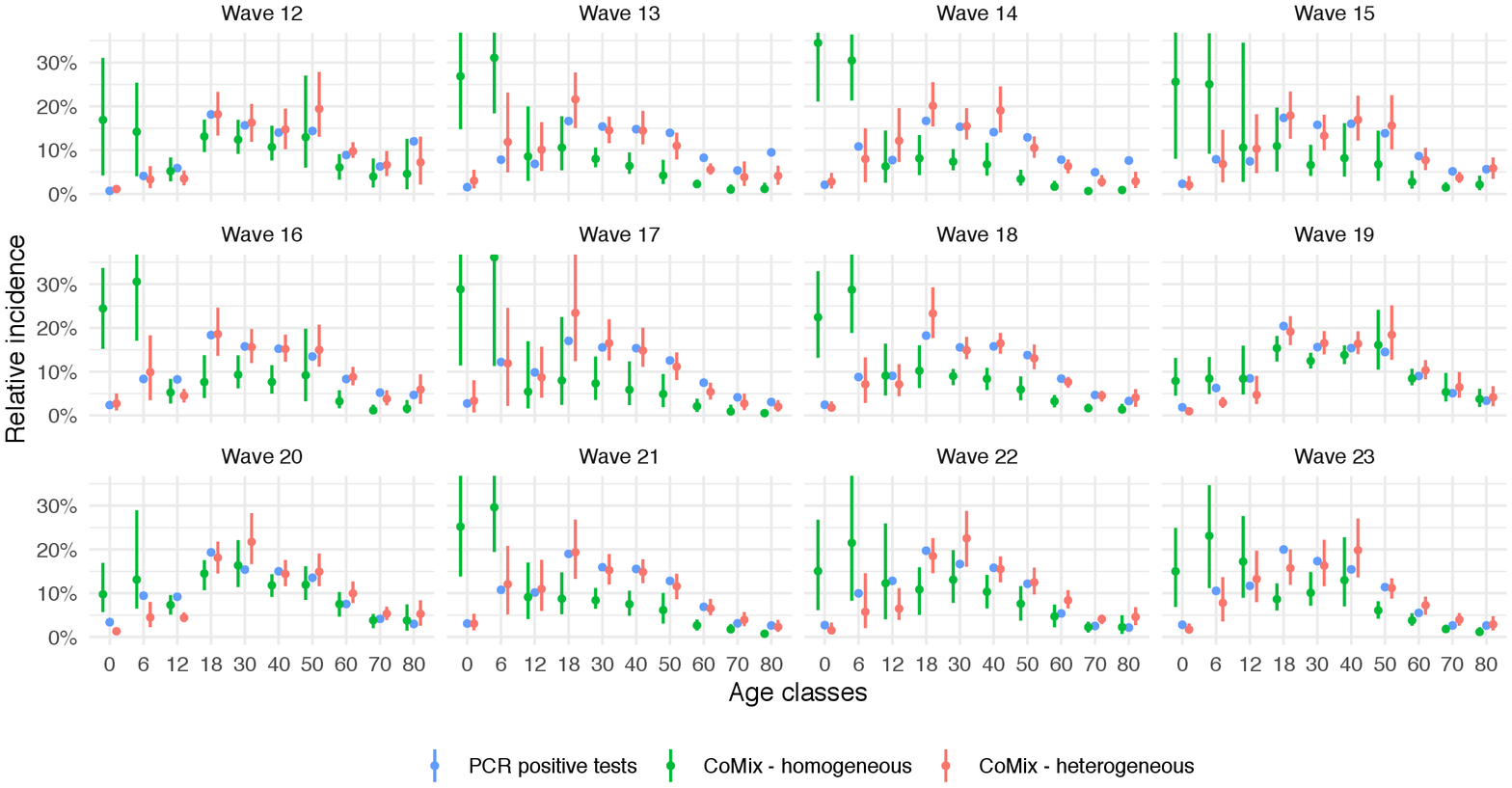}
\end{center}
\caption{Relative incidence using estimated $q$-susceptibility and assumption on infectiousness {\scriptsize$(1,1,1,1,1,1,1,1,1,1)$}.}
\label{sus1homong}
\end{figure}

\begin{figure}[htb!]
\begin{center}
\includegraphics[width=0.9\textwidth]{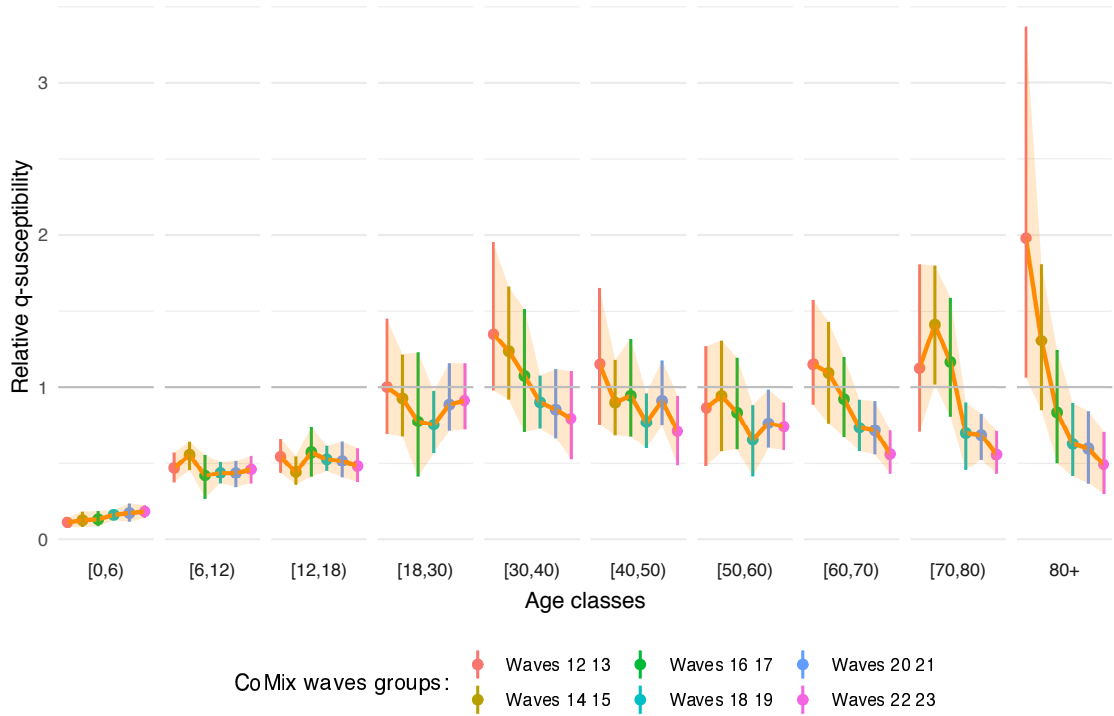}
\end{center}
\caption{Estimation of relative $q$-susceptibility with time evolution using assumption on infectiousness {\scriptsize$(1,1,1,1,1,1,1,1,1,1)$}.}
\label{sus6homo}
\end{figure}

\begin{figure}[htb!]
\begin{center}
\includegraphics[width=0.9\textwidth]{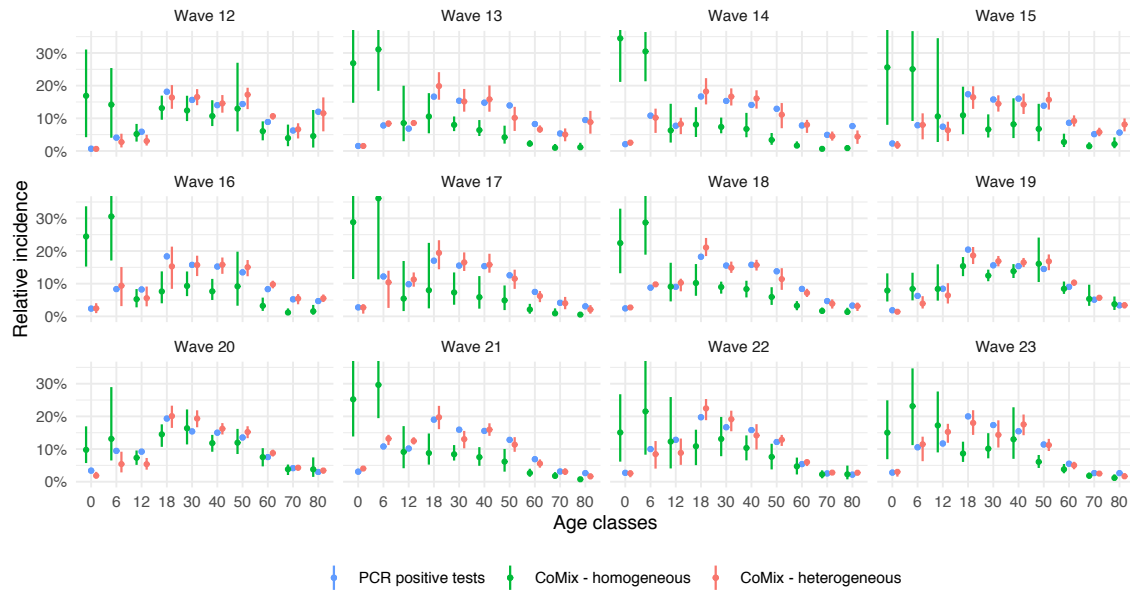}
\end{center}
\caption{Relative incidence using estimated $q$-susceptibility with time evolution and assumption on infectiousness {\scriptsize$(1,1,1,1,1,1,1,1,1,1)$}.}
\label{sus6homong}
\end{figure}

\begin{figure}[htb!]
\begin{center}
\includegraphics[width=0.5\textwidth]{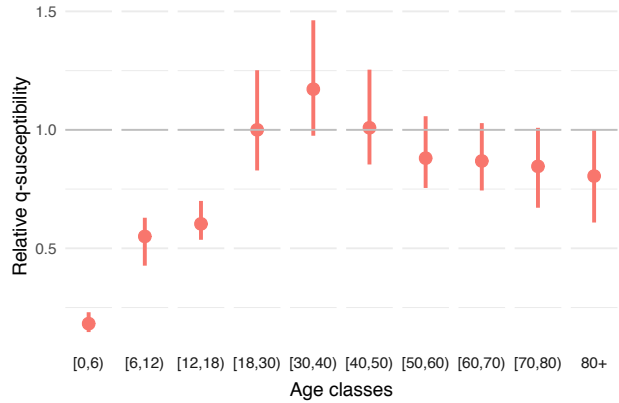}
\end{center}
\caption{Estimation of relative $q$-susceptibility using assumption on infectiousness {\scriptsize$(0.54,0.55,0.56,0.59,0.7,0.76,0.9,0.99,0.99,0.99)$}.}
\label{sus1ass}
\end{figure}

\begin{figure}[htb!]
\begin{center}
\includegraphics[width=0.9\textwidth]{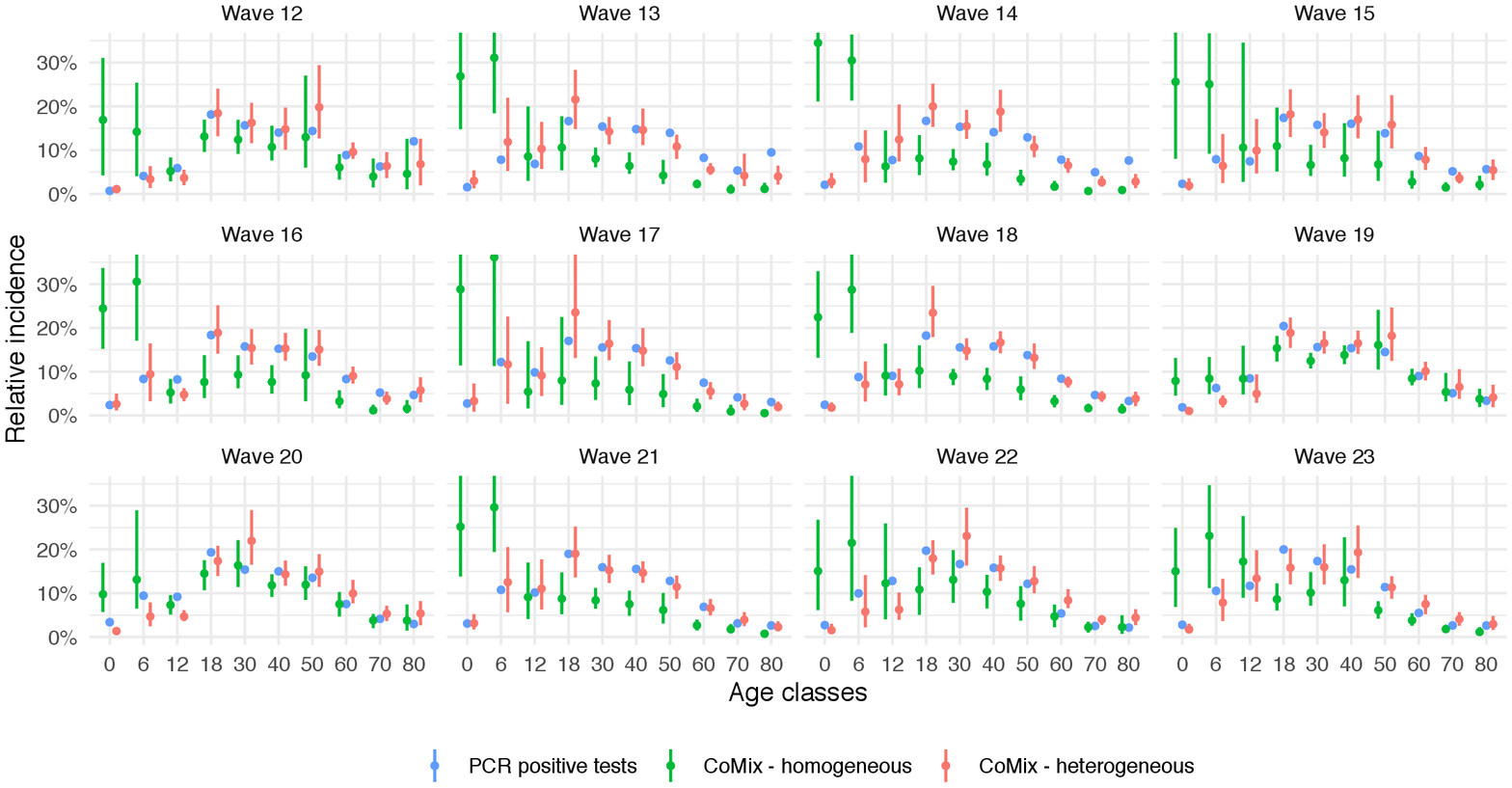}
\end{center}
\caption{Relative incidence using estimated $q$-susceptibility and assumption on infectiousness {\scriptsize$(0.54,0.55,0.56,0.59,0.7,0.76,0.9,0.99,0.99,0.99)$}.}
\label{sus1assng}
\end{figure}

\begin{figure}[htb!]
\begin{center}
\includegraphics[width=0.9\textwidth]{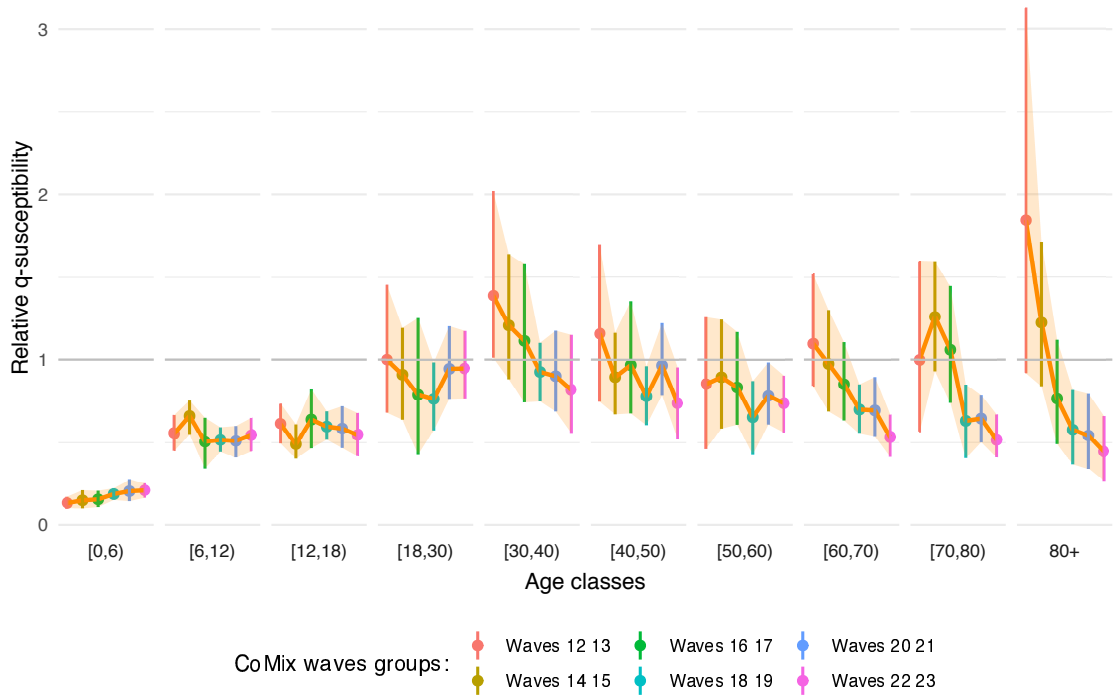}
\end{center}
\caption{Estimation of relative $q$-susceptibility with time evolution using assumption on infectiousness {\scriptsize$(0.54,0.55,0.56,0.59,0.7,0.76,0.9,0.99,0.99,0.99)$}.}
\label{sus6ass}
\end{figure}

\begin{figure}[htb!]
\begin{center}
\includegraphics[width=0.9\textwidth]{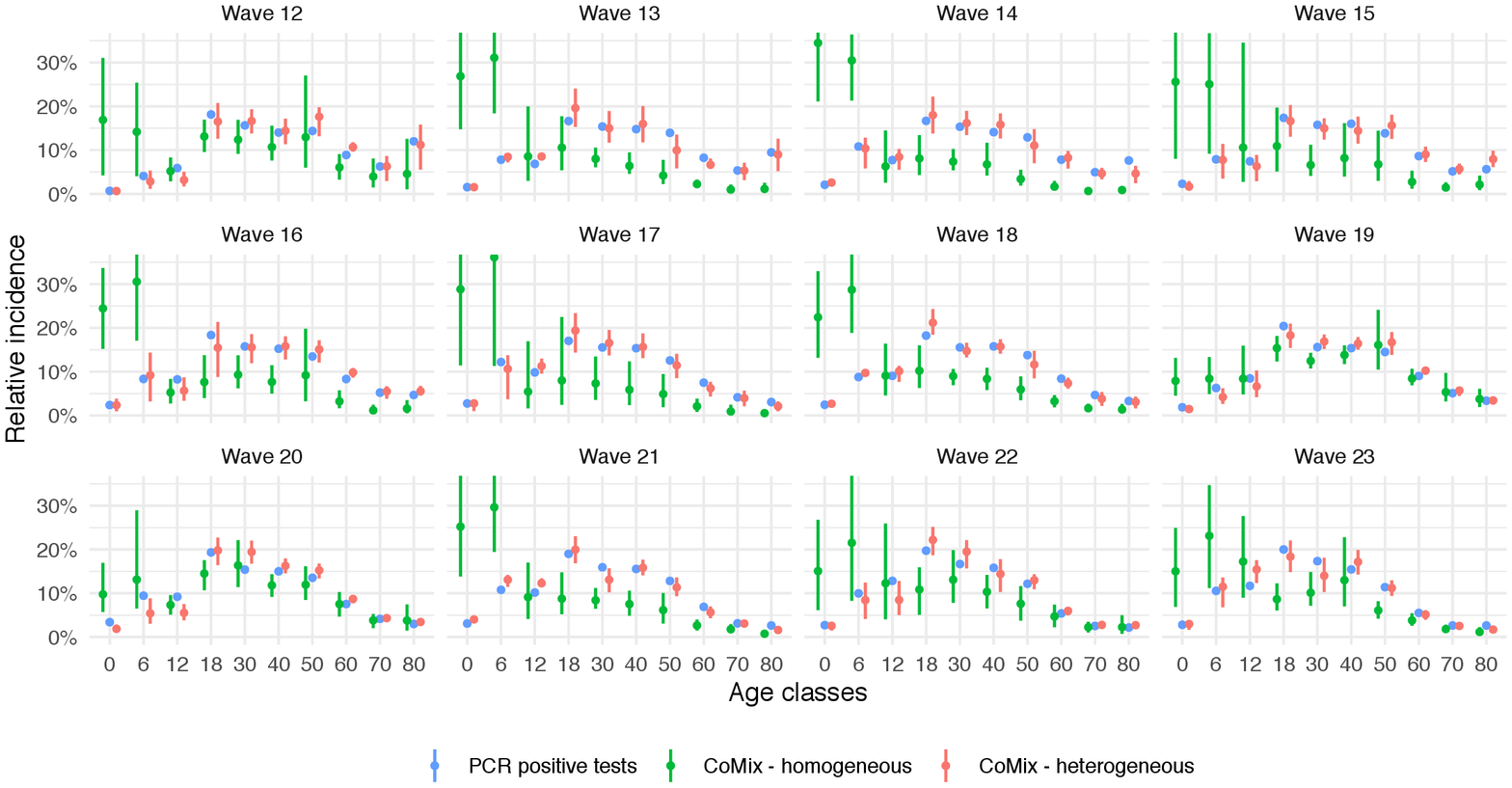}
\end{center}
\caption{Relative incidence using estimated $q$-susceptibility with time evolution and assumption on infectiousness {\scriptsize$(0.54,0.55,0.56,0.59,0.7,0.76,0.9,0.99,0.99,0.99)$}.}
\label{sus6assng}
\end{figure}

\begin{figure}[htb!]
\begin{center}
\includegraphics[width=0.5\textwidth]{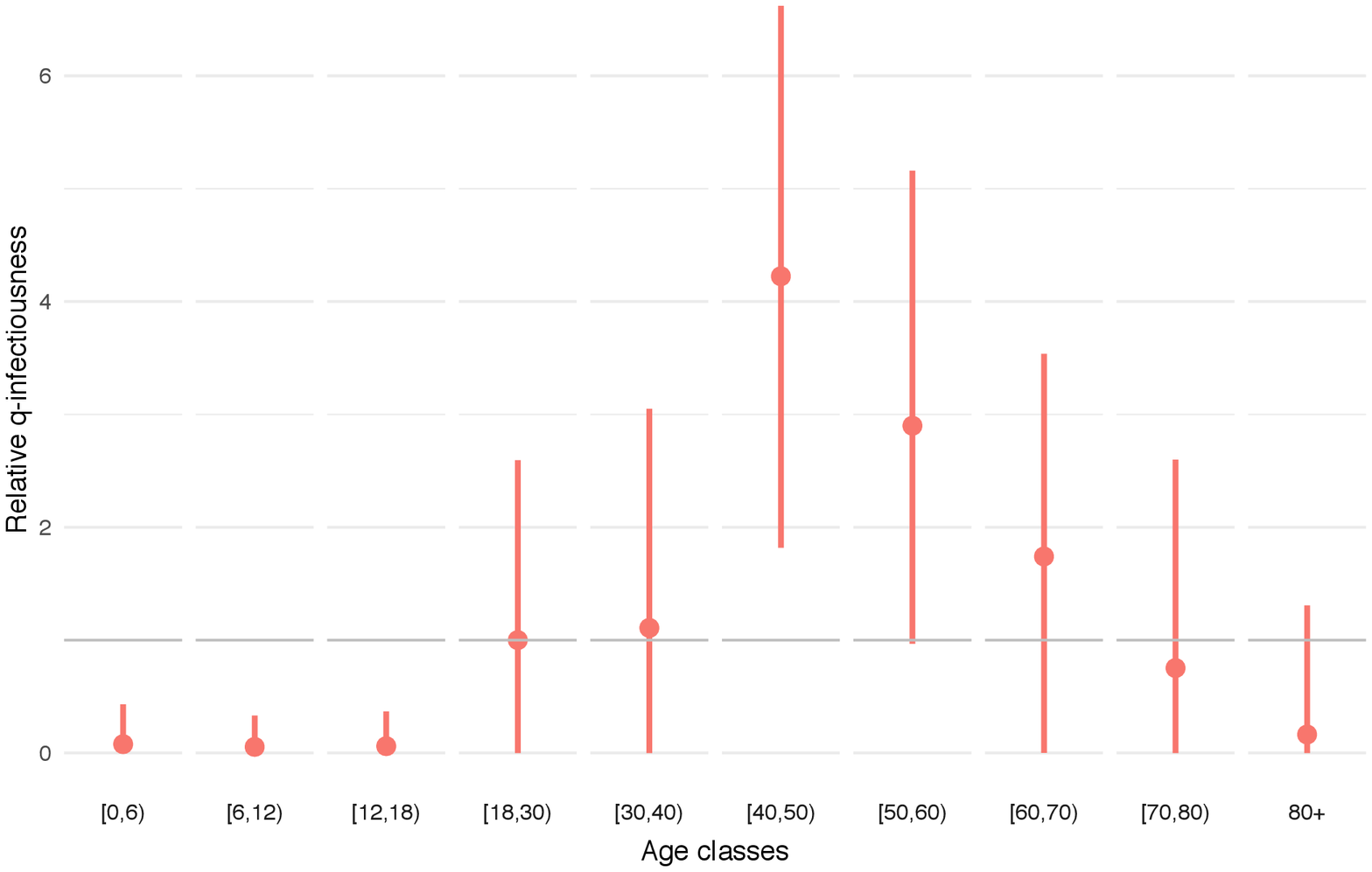}
\end{center}
\caption{Estimation of relative $q$-infectiousness using assumption on susceptibility {\scriptsize$(1,1,1,1,1,1,1,1,1,1)$}.}
\label{inf1homo}
\end{figure}

\begin{figure}[htb!]
\begin{center}
\includegraphics[width=0.9\textwidth]{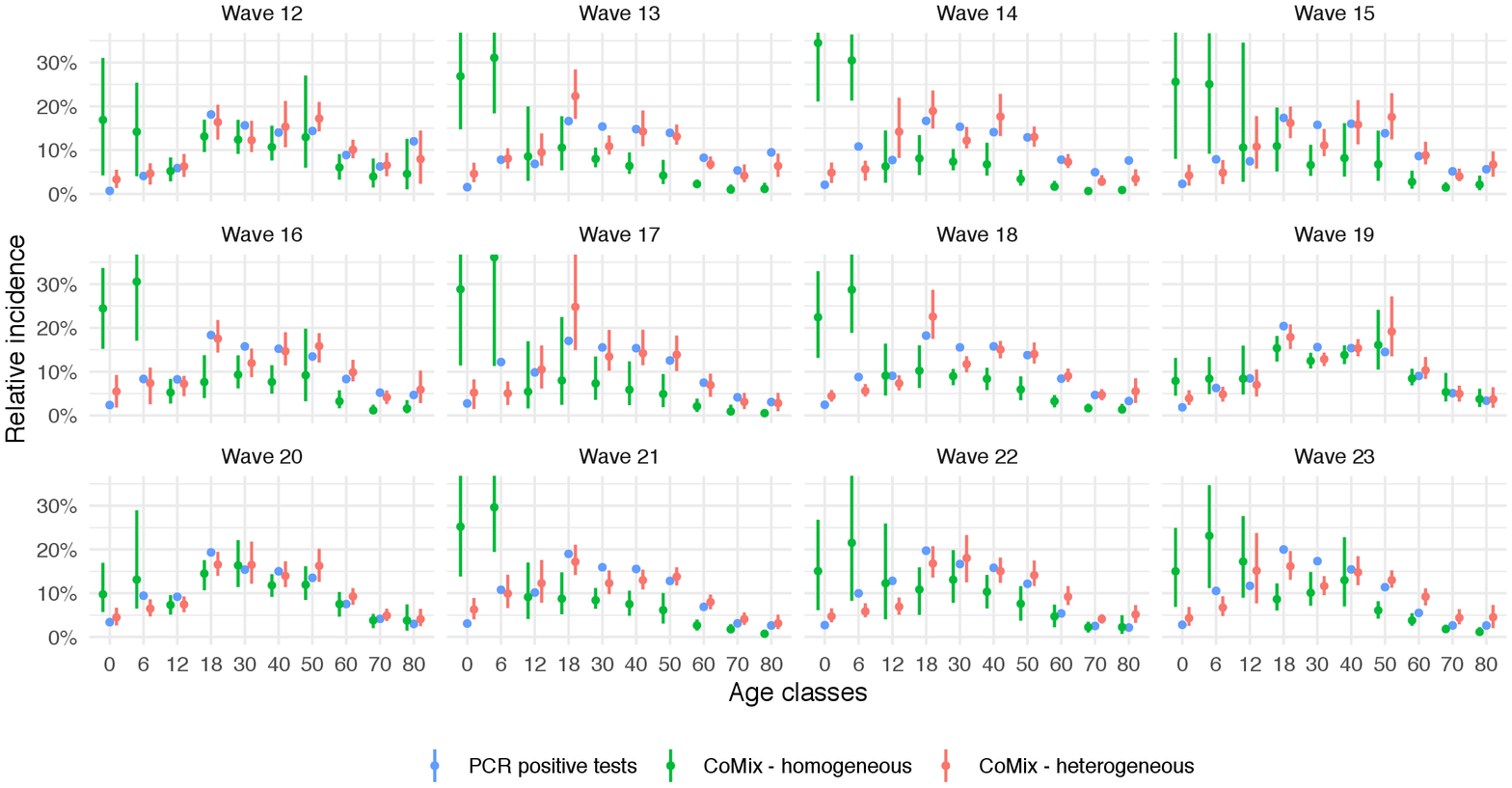}
\end{center}
\caption{Relative incidence using estimated $q$-infectiousness and assumption on susceptibility {\scriptsize$(1,1,1,1,1,1,1,1,1,1)$}.}
\label{inf1homong}
\end{figure}

\begin{figure}[htb!]
\begin{center}
\includegraphics[width=0.9\textwidth]{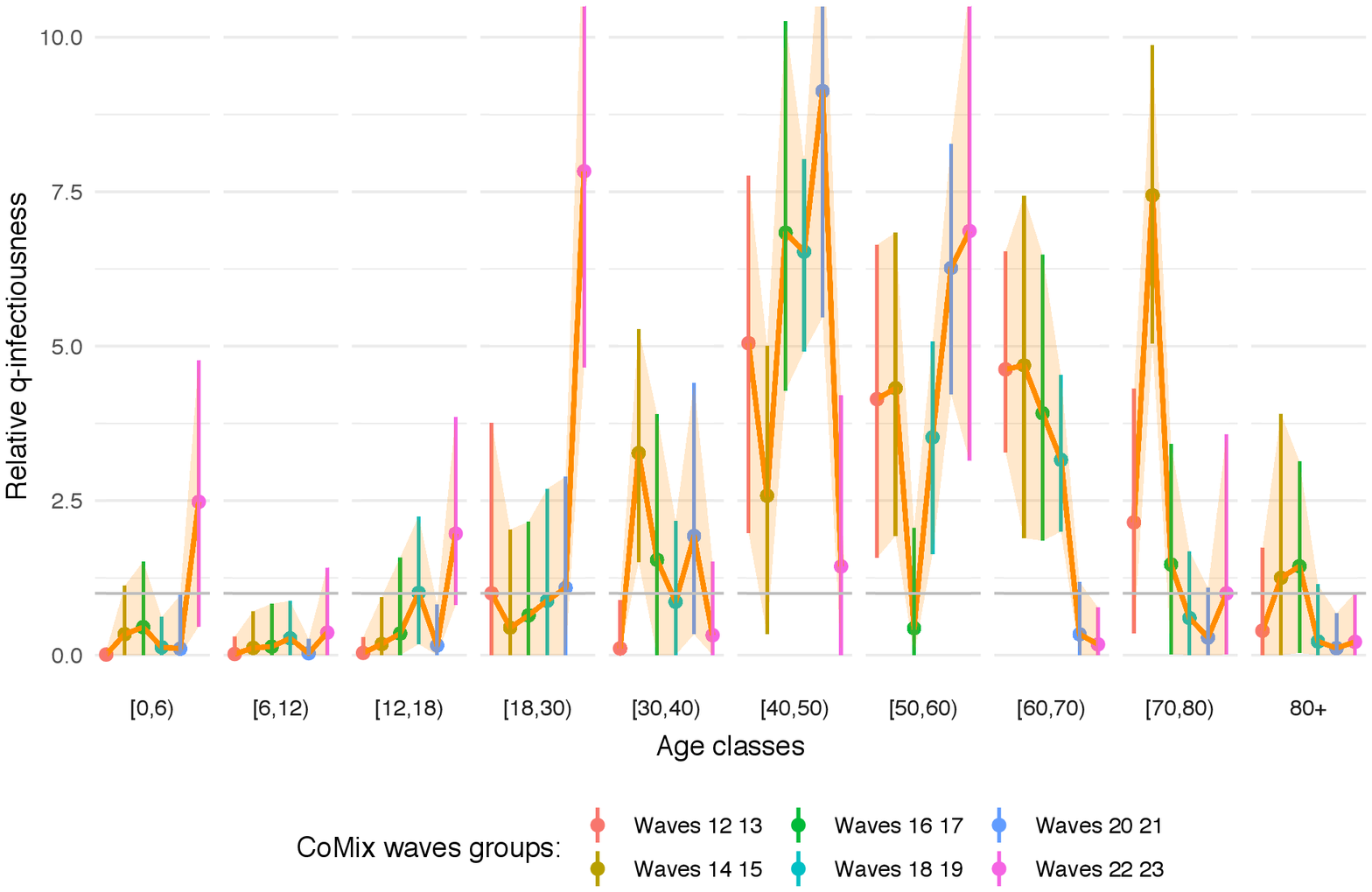}
\end{center}
\caption{Estimation of relative $q$-infectiousness with time evolution using assumption on susceptibility {\scriptsize$(1,1,1,1,1,1,1,1,1,1)$}.}
\label{inf6homo}
\end{figure}

\begin{figure}[htb!]
\begin{center}
\includegraphics[width=0.9\textwidth]{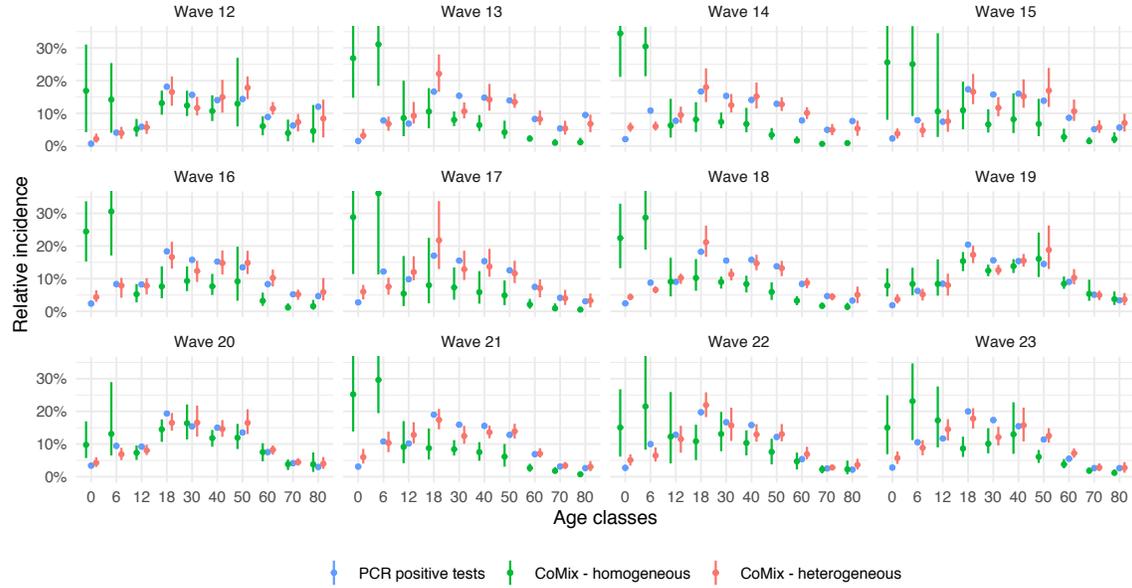}
\end{center}
\caption{Relative incidence using estimated $q$-infectiousness with time evolution and assumption on susceptibility {\scriptsize$(1,1,1,1,1,1,1,1,1,1)$}.}
\label{inf6homong}
\end{figure}

\begin{figure}[htb!]
\begin{center}
\includegraphics[width=0.5\textwidth]{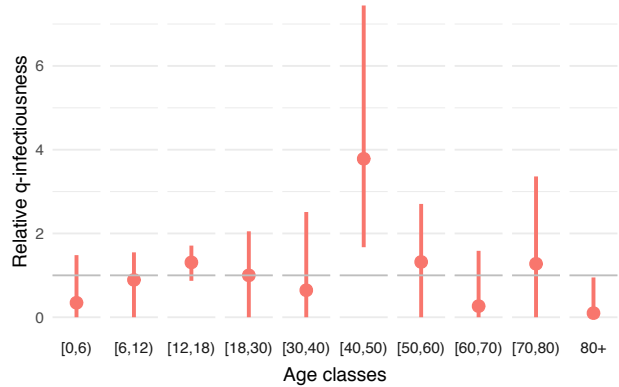}
\end{center}
\caption{Estimation of relative $q$-infectiousness using assumption on susceptibility {\scriptsize$(0.4,0.39,0.38,0.79,0.86,0.8,0.82,0.88,0.74,0.74)$}.}
\label{inf1ass}
\end{figure}

\begin{figure}[htb!]
\begin{center}
\includegraphics[width=0.9\textwidth]{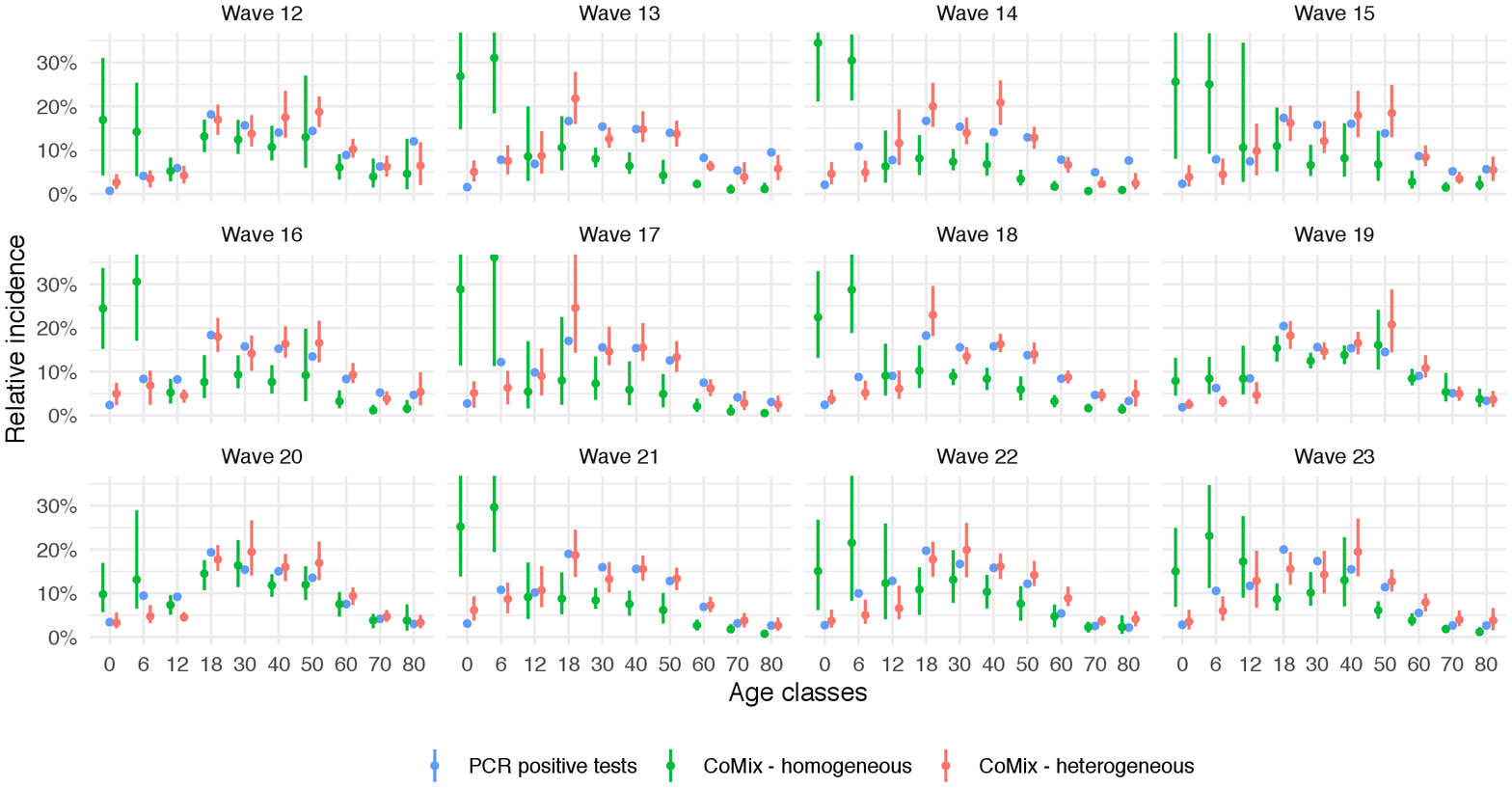}
\end{center}
\caption{Relative incidence using estimated $q$-infectiousness and assumption on susceptibility {\scriptsize$(0.4,0.39,0.38,0.79,0.86,0.8,0.82,0.88,0.74,0.74)$}.}
\label{inf1assng}
\end{figure}

\begin{figure}[htb!]
\begin{center}
\includegraphics[width=0.9\textwidth]{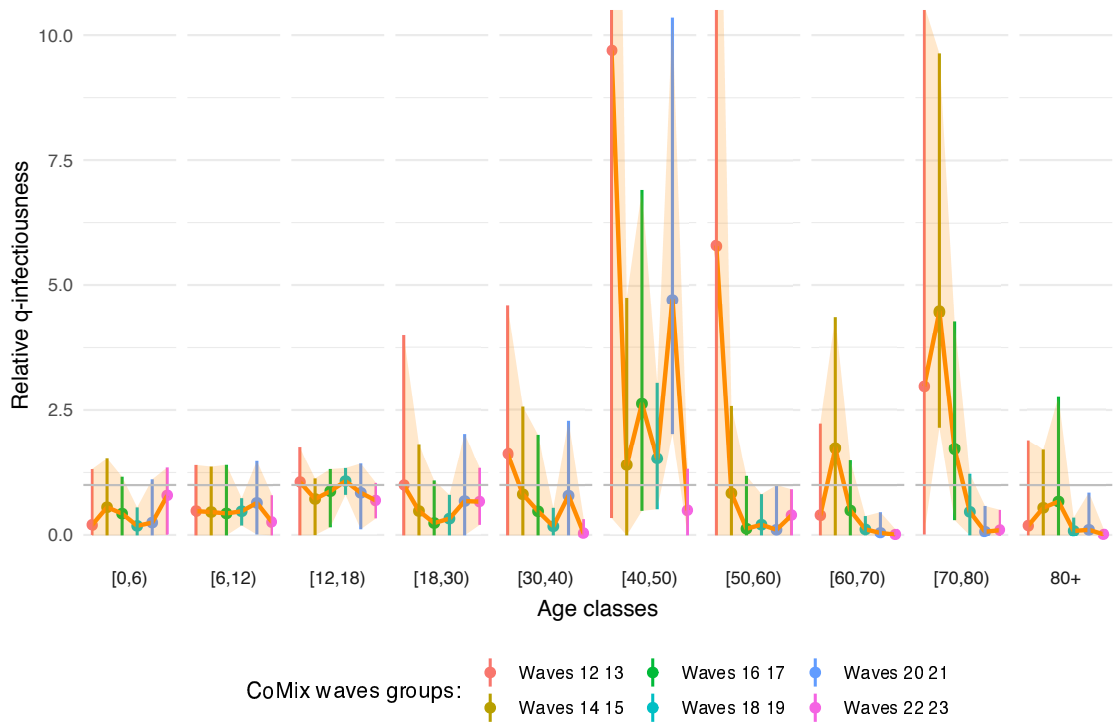}
\end{center}
\caption{Estimation of relative $q$-infectiousness with time evolution using assumption on susceptibility {\scriptsize$(0.4,0.39,0.38,0.79,0.86,0.8,0.82,0.88,0.74,0.74)$}.}
\label{inf6ass}
\end{figure}

\begin{figure}[htb!]
\begin{center}
\includegraphics[width=0.9\textwidth]{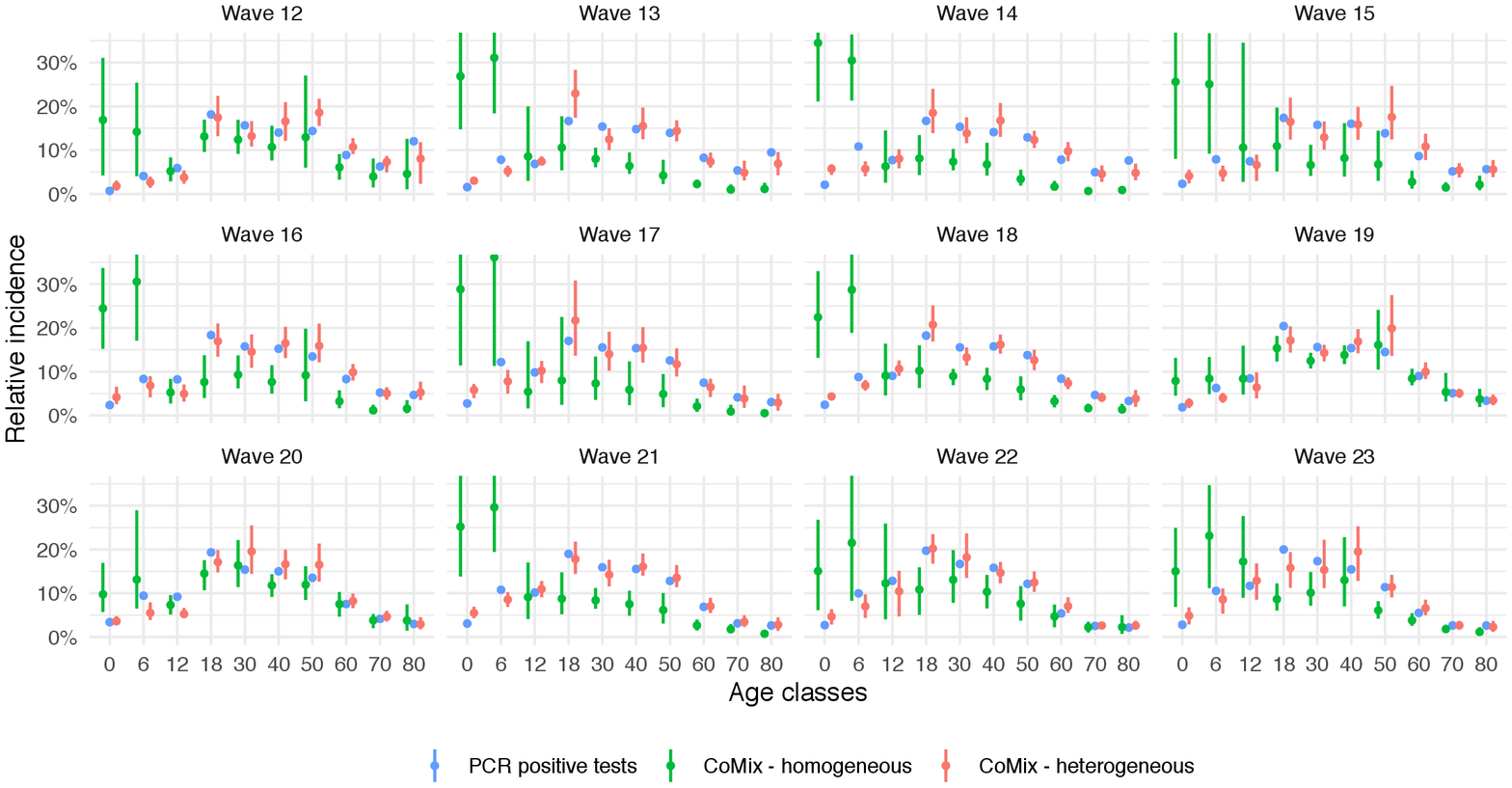}
\end{center}
\caption{Relative incidence using estimated $q$-infectiousness with time evolution and assumption on susceptibility {\scriptsize$(0.4,0.39,0.38,0.79,0.86,0.8,0.82,0.88,0.74,0.74)$}.}
\label{inf6assng}
\end{figure}

Since our method depends on assumptions made on $q$-susceptibility $(a_i)$ or $q$-infectiousness $(h_j)$ vectors, we present an additional sensitivity analysis concerning those assumptions. This analysis is performed by fixing a single bootstrap (hence removing the uncertainty related to social contact data) but varying the assumption on the $q$-infectiousness vector $(h_j)$ or the $q$-susceptibility vector $(a_j)$. We start with the heterogeneous infectiousness assumption $(h_j)=(0.54,0.55,0.56,0.59,0.7,0.76,0.9,0.99,0.99,0.99)$ or $(a_j)=(0.4,0.39,0.38,0.79,0.86,0.8,0.82,0.88,0.74,0.74)$ and perform a random uniform additive variation $\mathcal U{[-0.1,0.1]}$ on each of the 10 components. Figure \ref{sensitivity_1} represents the complete variation of the estimate on relative $q$-susceptibility $(a_i)$ (analysis similar to Fig~\ref{sus1ass}) under 200 simulated random variations on $(h_j)$ while Figure \ref{sensitivity_2} represents the complete variation of the estimate on relative $q$-susceptibility $(h_i)$ (analysis similar to Fig~\ref{inf1ass}) under 200 simulated random variations on $(a_j)$.

\clearpage

\subsection*{Estimation of $q$-susceptibility using homogeneous infectiousness}

Method: Estimation of the $(a_i)$ relative $q$-susceptibility vector.\\
Assumption: homogeneous infectiousness  $(h_j) = (1,1,1,1,1,1,1,1,1,1)$.\\
Normalization method: Mean $q$-susceptibility among children age classes [0,6),  [6,12) and [12,18) is assumed constant among bootstraps and wave groups (if applicable). The mean of the first adult age class [18,30) is set to 1 for the first period.\\

\begin{table}[ht]
\scriptsize
\centering
\begin{tabular}{lrrrrr}
  \hline
ageclass & mean & median & sd & lower & upper \\ 
  \hline
[0,6) & 0.157 & 0.155 & 0.020 & 0.124 & 0.204 \\\relax  
  [6,12) & 0.474 & 0.481 & 0.045 & 0.367 & 0.542 \\\relax  
  [12,18) & 0.540 & 0.535 & 0.039 & 0.478 & 0.619 \\\relax  
  [18,30) & 1.000 & 0.993 & 0.103 & 0.825 & 1.238 \\\relax  
  [30,40) & 1.161 & 1.149 & 0.122 & 0.973 & 1.443 \\\relax  
  [40,50) & 1.003 & 0.992 & 0.100 & 0.846 & 1.256 \\\relax  
  [50,60) & 0.897 & 0.889 & 0.083 & 0.765 & 1.063 \\\relax  
  [60,70) & 0.936 & 0.935 & 0.073 & 0.806 & 1.100 \\\relax  
  [70,80) & 0.953 & 0.951 & 0.095 & 0.768 & 1.150 \\ 
  80+ & 0.917 & 0.914 & 0.115 & 0.695 & 1.167 \\ 
   \hline
\end{tabular}
\caption{Relative $q$-susceptibility using assumption on infectiousness {\scriptsize$(1,1,1,1,1,1,1,1,1,1)$} corresponding to Figure \ref{sus1homo}.} 
\label{sus1homotable}
\end{table}

\begin{table}[ht]
\scriptsize
\begin{subtable}{0.48\textwidth}
\centering
\begin{tabular}{lrrrrr}
  \hline
ageclass & mean & median & sd & lower & upper \\ 
  \hline
[0,6) & 0.112 & 0.112 & 0.015 & 0.081 & 0.143 \\\relax 
  [6,12) & 0.469 & 0.470 & 0.053 & 0.374 & 0.571 \\\relax 
  [12,18) & 0.543 & 0.543 & 0.057 & 0.438 & 0.659 \\\relax 
  [18,30) & 1.000 & 0.966 & 0.201 & 0.691 & 1.449 \\\relax 
  [30,40) & 1.347 & 1.314 & 0.225 & 0.978 & 1.949 \\\relax 
  [40,50) & 1.150 & 1.114 & 0.240 & 0.755 & 1.649 \\\relax 
  [50,60) & 0.864 & 0.862 & 0.211 & 0.481 & 1.265 \\\relax 
  [60,70) & 1.149 & 1.114 & 0.178 & 0.885 & 1.573 \\\relax 
  [70,80) & 1.123 & 1.113 & 0.258 & 0.706 & 1.806 \\ 
  80+ & 1.979 & 1.885 & 0.583 & 1.062 & 3.367 \\ 
   \hline
\end{tabular}
\caption{Waves 12 13} 
\end{subtable}
\qquad
\begin{subtable}{0.48\textwidth}
\centering
\begin{tabular}{lrrrrr}
  \hline
ageclass & mean & median & sd & lower & upper \\ 
  \hline
[0,6) & 0.128 & 0.125 & 0.024 & 0.084 & 0.183 \\\relax 
  [6,12) & 0.555 & 0.561 & 0.048 & 0.455 & 0.636 \\\relax 
  [12,18) & 0.441 & 0.437 & 0.051 & 0.360 & 0.548 \\\relax 
  [18,30) & 0.926 & 0.912 & 0.137 & 0.675 & 1.210 \\\relax 
  [30,40) & 1.232 & 1.219 & 0.174 & 0.918 & 1.657 \\\relax 
  [40,50) & 0.899 & 0.890 & 0.124 & 0.683 & 1.176 \\\relax 
  [50,60) & 0.942 & 0.949 & 0.183 & 0.579 & 1.300 \\\relax 
  [60,70) & 1.091 & 1.087 & 0.171 & 0.762 & 1.427 \\\relax 
  [70,80) & 1.409 & 1.393 & 0.199 & 1.016 & 1.797 \\ 
  80+ & 1.303 & 1.286 & 0.238 & 0.850 & 1.806 \\ 
   \hline
\end{tabular}
\caption{Waves 14 15} 
\end{subtable}
\par\bigskip
\begin{subtable}{0.48\textwidth}
\centering
\begin{tabular}{lrrrrr}
  \hline
ageclass & mean & median & sd & lower & upper \\ 
  \hline
[0,6) & 0.133 & 0.131 & 0.025 & 0.089 & 0.186 \\\relax 
  [6,12) & 0.419 & 0.424 & 0.073 & 0.270 & 0.556 \\\relax 
  [12,18) & 0.572 & 0.568 & 0.084 & 0.409 & 0.735 \\\relax 
  [18,30) & 0.775 & 0.761 & 0.218 & 0.410 & 1.225 \\\relax 
  [30,40) & 1.072 & 1.053 & 0.217 & 0.704 & 1.516 \\\relax 
  [40,50) & 0.944 & 0.926 & 0.169 & 0.675 & 1.314 \\\relax 
  [50,60) & 0.833 & 0.826 & 0.153 & 0.591 & 1.189 \\\relax 
  [60,70) & 0.921 & 0.917 & 0.133 & 0.671 & 1.195 \\\relax 
  [70,80) & 1.162 & 1.149 & 0.202 & 0.807 & 1.592 \\ 
  80+ & 0.835 & 0.826 & 0.194 & 0.501 & 1.240 \\ 
   \hline
\end{tabular}
\caption{Waves 16 17} 
\end{subtable}
\qquad
\begin{subtable}{0.48\textwidth}
\centering
\begin{tabular}{lrrrrr}
  \hline
ageclass & mean & median & sd & lower & upper \\ 
  \hline
[0,6) & 0.162 & 0.161 & 0.018 & 0.126 & 0.194 \\\relax 
  [6,12) & 0.435 & 0.437 & 0.035 & 0.368 & 0.509 \\\relax 
  [12,18) & 0.527 & 0.526 & 0.041 & 0.451 & 0.612 \\\relax 
  [18,30) & 0.753 & 0.756 & 0.111 & 0.568 & 0.974 \\\relax 
  [30,40) & 0.901 & 0.898 & 0.098 & 0.726 & 1.074 \\\relax 
  [40,50) & 0.769 & 0.761 & 0.095 & 0.599 & 0.958 \\\relax 
  [50,60) & 0.652 & 0.643 & 0.129 & 0.415 & 0.883 \\\relax 
  [60,70) & 0.734 & 0.731 & 0.087 & 0.580 & 0.918 \\\relax 
  [70,80) & 0.697 & 0.694 & 0.125 & 0.456 & 0.900 \\ 
  80+ & 0.626 & 0.613 & 0.129 & 0.413 & 0.897 \\ 
   \hline
\end{tabular}
\caption{Waves 18 19} 
\end{subtable}
\par\bigskip
\begin{subtable}{0.48\textwidth}
\centering
\begin{tabular}{lrrrrr}
  \hline
ageclass & mean & median & sd & lower & upper \\ 
  \hline
[0,6) & 0.173 & 0.172 & 0.030 & 0.117 & 0.235 \\\relax 
  [6,12) & 0.434 & 0.434 & 0.049 & 0.339 & 0.515 \\\relax 
  [12,18) & 0.517 & 0.517 & 0.063 & 0.406 & 0.638 \\\relax 
  [18,30) & 0.886 & 0.869 & 0.120 & 0.712 & 1.154 \\\relax 
  [30,40) & 0.853 & 0.839 & 0.121 & 0.663 & 1.117 \\\relax 
  [40,50) & 0.911 & 0.904 & 0.112 & 0.747 & 1.172 \\\relax 
  [50,60) & 0.762 & 0.754 & 0.104 & 0.601 & 0.983 \\\relax 
  [60,70) & 0.715 & 0.712 & 0.091 & 0.560 & 0.909 \\\relax 
  [70,80) & 0.683 & 0.683 & 0.078 & 0.525 & 0.826 \\ 
  80+ & 0.595 & 0.601 & 0.126 & 0.366 & 0.843 \\ 
   \hline
\end{tabular}
\caption{Waves 20 21} 
\end{subtable}
\qquad
\begin{subtable}{0.48\textwidth}
\centering
\begin{tabular}{lrrrrr}
  \hline
ageclass & mean & median & sd & lower & upper \\ 
  \hline
[0,6) & 0.183 & 0.182 & 0.021 & 0.141 & 0.224 \\\relax 
  [6,12) & 0.460 & 0.460 & 0.047 & 0.367 & 0.550 \\\relax 
  [12,18) & 0.482 & 0.482 & 0.058 & 0.376 & 0.597 \\\relax 
  [18,30) & 0.912 & 0.904 & 0.112 & 0.720 & 1.154 \\\relax 
  [30,40) & 0.794 & 0.784 & 0.153 & 0.530 & 1.103 \\\relax 
  [40,50) & 0.710 & 0.710 & 0.125 & 0.490 & 0.943 \\\relax 
  [50,60) & 0.740 & 0.745 & 0.088 & 0.586 & 0.899 \\\relax 
  [60,70) & 0.561 & 0.557 & 0.070 & 0.433 & 0.715 \\\relax 
  [70,80) & 0.557 & 0.553 & 0.074 & 0.432 & 0.712 \\ 
  80+ & 0.492 & 0.486 & 0.101 & 0.297 & 0.704 \\ 
  \hline
\end{tabular}
\caption{Waves 22 23} 
\end{subtable}
\caption{Relative $q$-susceptibility with time evolution using assumption on infectiousness {\scriptsize$(1,1,1,1,1,1,1,1,1,1)$} corresponding to Figure \ref{sus6homo}.} 
\label{sus6homotable}
\end{table}

\clearpage
\subsection*{Estimation of $q$-susceptibility using heterogeneous infectiousness}

Method: Estimation of the $(a_i)$ relative $q$-susceptibility vector.\\
Assumption: heterogeneous infectiousness  $(h_j) = (0.54,0.55,0.56,0.59,0.7,0.76,0.9,0.99,0.99,0.99)$ using the proportion of asymptomatic cases in the Belgian population with asymptomatic infectiousness assumed at 0.51 as used in \cite{ABRAMS2021100449} using data from \cite{Wu:2020tk}.\\
Normalization method: Mean $q$-susceptibility among children age classes [0,6),  [6,12) and [12,18) is assumed constant among bootstraps and wave groups (if applicable). The mean of the first adult age class [18,30) is set to 1 for the first period.

\begin{table}[ht]
\scriptsize
\centering
\begin{tabular}{lrrrrr}
  \hline
ageclass & mean & median & sd & lower & upper \\ 
  \hline
[0,6) & 0.182 & 0.180 & 0.020 & 0.146 & 0.230 \\\relax 
  [6,12) & 0.550 & 0.559 & 0.051 & 0.427 & 0.629 \\\relax 
  [12,18) & 0.603 & 0.599 & 0.044 & 0.536 & 0.700 \\\relax 
  [18,30) & 1.000 & 0.993 & 0.102 & 0.829 & 1.252 \\\relax 
  [30,40) & 1.172 & 1.162 & 0.120 & 0.975 & 1.462 \\\relax 
  [40,50) & 1.009 & 0.997 & 0.100 & 0.854 & 1.254 \\\relax 
  [50,60) & 0.880 & 0.875 & 0.082 & 0.755 & 1.057 \\\relax
  [60,70) & 0.869 & 0.864 & 0.070 & 0.744 & 1.028 \\\relax 
  [70,80) & 0.846 & 0.849 & 0.090 & 0.671 & 1.009 \\\relax 
  80+ & 0.805 & 0.798 & 0.101 & 0.609 & 1.000 \\ 
   \hline
\end{tabular}
\caption{Relative $q$-susceptibility using assumption on infectiousness {\scriptsize$(0.54,0.55,0.56,0.59,0.7,0.76,0.9,0.99,0.99,0.99)$} corresponding to Figure \ref{sus1ass}.} 
\label{sus1asstable}
\end{table}

\begin{table}[ht]
\scriptsize
\begin{subtable}{0.48\textwidth}
\centering
\begin{tabular}{lrrrrr}
  \hline
ageclass & mean & median & sd & lower & upper \\ 
  \hline
[0,6) & 0.134 & 0.134 & 0.018 & 0.103 & 0.172 \\\relax 
  [6,12) & 0.555 & 0.556 & 0.059 & 0.449 & 0.666 \\\relax 
  [12,18) & 0.613 & 0.616 & 0.064 & 0.495 & 0.735 \\\relax 
  [18,30) & 1.000 & 0.971 & 0.208 & 0.680 & 1.454 \\\relax 
  [30,40) & 1.387 & 1.346 & 0.238 & 1.012 & 2.019 \\\relax 
  [40,50) & 1.157 & 1.125 & 0.249 & 0.746 & 1.695 \\\relax 
  [50,60) & 0.853 & 0.853 & 0.216 & 0.461 & 1.255 \\\relax 
  [60,70) & 1.096 & 1.066 & 0.169 & 0.839 & 1.519 \\\relax 
  [70,80) & 0.999 & 0.976 & 0.249 & 0.563 & 1.593 \\ 
  80+ & 1.844 & 1.763 & 0.567 & 0.920 & 3.127 \\ 
   \hline
\end{tabular}
\caption{Waves 12 13} 
\end{subtable}
\qquad
\begin{subtable}{0.48\textwidth}
\centering
\begin{tabular}{lrrrrr}
  \hline
ageclass & mean & median & sd & lower & upper \\ 
  \hline
[0,6) & 0.151 & 0.148 & 0.027 & 0.103 & 0.212 \\\relax 
  [6,12) & 0.660 & 0.663 & 0.056 & 0.548 & 0.753 \\\relax 
  [12,18) & 0.491 & 0.484 & 0.058 & 0.402 & 0.611 \\\relax 
  [18,30) & 0.911 & 0.912 & 0.136 & 0.639 & 1.191 \\\relax 
  [30,40) & 1.205 & 1.189 & 0.178 & 0.883 & 1.637 \\\relax 
  [40,50) & 0.896 & 0.888 & 0.123 & 0.670 & 1.161 \\\relax 
  [50,60) & 0.895 & 0.893 & 0.177 & 0.585 & 1.241 \\\relax 
  [60,70) & 0.975 & 0.980 & 0.155 & 0.687 & 1.299 \\\relax 
  [70,80) & 1.258 & 1.257 & 0.181 & 0.931 & 1.590 \\ 
  80+ & 1.222 & 1.198 & 0.231 & 0.841 & 1.715 \\ 
   \hline
\end{tabular}
\caption{Waves 14 15} 
\end{subtable}
\par\bigskip
\begin{subtable}{0.48\textwidth}
\centering
\begin{tabular}{lrrrrr}
  \hline
ageclass & mean & median & sd & lower & upper \\ 
  \hline
[0,6) & 0.157 & 0.157 & 0.026 & 0.110 & 0.209 \\\relax 
  [6,12) & 0.506 & 0.513 & 0.081 & 0.341 & 0.650 \\\relax 
  [12,18) & 0.639 & 0.630 & 0.093 & 0.466 & 0.822 \\\relax 
  [18,30) & 0.789 & 0.785 & 0.221 & 0.428 & 1.250 \\\relax 
  [30,40) & 1.113 & 1.089 & 0.222 & 0.742 & 1.577 \\\relax 
  [40,50) & 0.968 & 0.957 & 0.170 & 0.676 & 1.353 \\\relax 
  [50,60) & 0.834 & 0.825 & 0.148 & 0.609 & 1.166 \\\relax 
  [60,70) & 0.853 & 0.854 & 0.122 & 0.634 & 1.105 \\\relax 
  [70,80) & 1.060 & 1.041 & 0.187 & 0.739 & 1.446 \\ 
  80+ & 0.765 & 0.746 & 0.159 & 0.491 & 1.119 \\ 
   \hline
\end{tabular}
\caption{Waves 16 17} 
\end{subtable}
\qquad
\begin{subtable}{0.48\textwidth}
\centering
\begin{tabular}{lrrrrr}
  \hline
ageclass & mean & median & sd & lower & upper \\ 
  \hline
[0,6) & 0.189 & 0.188 & 0.018 & 0.153 & 0.224 \\\relax 
  [6,12) & 0.516 & 0.520 & 0.039 & 0.444 & 0.593 \\\relax 
  [12,18) & 0.597 & 0.593 & 0.044 & 0.517 & 0.688 \\\relax 
  [18,30) & 0.763 & 0.761 & 0.113 & 0.574 & 0.984 \\\relax 
  [30,40) & 0.927 & 0.925 & 0.097 & 0.748 & 1.101 \\\relax 
  [40,50) & 0.779 & 0.775 & 0.093 & 0.606 & 0.961 \\\relax 
  [50,60) & 0.654 & 0.645 & 0.123 & 0.428 & 0.872 \\\relax 
  [60,70) & 0.699 & 0.701 & 0.078 & 0.555 & 0.852 \\\relax 
  [70,80) & 0.632 & 0.640 & 0.122 & 0.405 & 0.851 \\ 
  80+ & 0.578 & 0.577 & 0.122 & 0.365 & 0.819 \\ 
   \hline
\end{tabular}
\caption{Waves 18 19} 
\end{subtable}
\par\bigskip
\begin{subtable}{0.48\textwidth}
\centering
\begin{tabular}{lrrrrr}
  \hline
ageclass & mean & median & sd & lower & upper \\ 
  \hline
[0,6) & 0.206 & 0.204 & 0.033 & 0.146 & 0.274 \\\relax 
  [6,12) & 0.511 & 0.513 & 0.055 & 0.407 & 0.601 \\\relax 
  [12,18) & 0.585 & 0.588 & 0.069 & 0.467 & 0.719 \\\relax 
  [18,30) & 0.946 & 0.936 & 0.119 & 0.763 & 1.201 \\\relax 
  [30,40) & 0.902 & 0.900 & 0.129 & 0.688 & 1.174 \\\relax 
  [40,50) & 0.965 & 0.953 & 0.115 & 0.785 & 1.219 \\\relax 
  [50,60) & 0.782 & 0.776 & 0.102 & 0.610 & 0.984 \\\relax 
  [60,70) & 0.694 & 0.691 & 0.091 & 0.537 & 0.896 \\\relax 
  [70,80) & 0.645 & 0.651 & 0.077 & 0.502 & 0.787 \\ 
  80+ & 0.543 & 0.530 & 0.122 & 0.338 & 0.796 \\ 
   \hline
\end{tabular}
\caption{Waves 20 21} 
\end{subtable}
\qquad
\begin{subtable}{0.48\textwidth}
\centering
\begin{tabular}{lrrrrr}
  \hline
ageclass & mean & median & sd & lower & upper \\ 
  \hline
[0,6) & 0.211 & 0.210 & 0.023 & 0.167 & 0.255 \\\relax 
  [6,12) & 0.545 & 0.545 & 0.054 & 0.447 & 0.649 \\\relax 
  [12,18) & 0.546 & 0.546 & 0.066 & 0.417 & 0.678 \\\relax 
  [18,30) & 0.948 & 0.929 & 0.111 & 0.766 & 1.172 \\\relax 
  [30,40) & 0.819 & 0.810 & 0.153 & 0.559 & 1.149 \\\relax 
  [40,50) & 0.737 & 0.735 & 0.122 & 0.524 & 0.954 \\\relax 
  [50,60) & 0.737 & 0.736 & 0.088 & 0.562 & 0.905 \\\relax 
  [60,70) & 0.532 & 0.530 & 0.067 & 0.414 & 0.668 \\\relax 
  [70,80) & 0.516 & 0.507 & 0.071 & 0.410 & 0.669 \\ 
  80+ & 0.446 & 0.439 & 0.099 & 0.266 & 0.660 \\ 
   \hline
\end{tabular}
\caption{Waves 22 23} 
\end{subtable}
\caption{Relative $q$-susceptibility with time evolution using assumption on infectiousness {\scriptsize$(0.54,0.55,0.56,0.59,0.7,0.76,0.9,0.99,0.99,0.99)$} corresponding to Figure \ref{sus6ass}.} 
\label{sus6asstable}
\end{table}

\clearpage
\subsection*{Estimation of $q$-infectiousness using homogeneous susceptibility}

Method: Estimation of the $(h_j)$ relative $q$-infectiousness vector.\\
Assumption: homogeneous susceptibility  $(a_i) = (1,1,1,1,1,1,1,1,1,1)$.\\
Normalization method: No normalization accross bootstraps and wave groups is applied here since the $q$-infectiousness among children age classes is estimated at $(0,0,0)$ for several bootstraps and this prevents using the same normaliztion method than in other subsections. The mean of the first adult age class [18,30) is set to 1 for the first period.

\begin{table}[ht]
\scriptsize
\centering
\begin{tabular}{lrrrrr}
  \hline
ageclass & mean & median & sd & lower & upper \\ 
  \hline
[0,6) & 0.078 & 0.000 & 0.129 & 0.000 & 0.430 \\\relax 
  [6,12) & 0.054 & 0.000 & 0.111 & 0.000 & 0.331 \\\relax 
  [12,18) & 0.060 & 0.000 & 0.106 & 0.000 & 0.368 \\\relax 
  [18,30) & 1.000 & 0.872 & 0.733 & 0.000 & 2.594 \\\relax 
  [30,40) & 1.108 & 0.991 & 0.910 & 0.000 & 3.049 \\\relax 
  [40,50) & 4.224 & 4.220 & 1.176 & 1.818 & 6.619 \\\relax 
  [50,60) & 2.899 & 2.896 & 1.144 & 0.968 & 5.159 \\\relax 
  [60,70) & 1.741 & 1.694 & 0.882 & 0.000 & 3.536 \\\relax 
  [70,80) & 0.752 & 0.513 & 0.810 & 0.000 & 2.598 \\ 
  80+ & 0.164 & 0.000 & 0.372 & 0.000 & 1.307 \\ 
   \hline
\end{tabular}
\caption{Relative $q$-infectiousness using assumption on susceptibility {\scriptsize$(1,1,1,1,1,1,1,1,1,1)$} corresponding to Figure \ref{inf1homo}.} 
\label{inf1homotable}
\end{table}

\begin{table}[ht]
\scriptsize
\begin{subtable}{0.48\textwidth}
\centering
\begin{tabular}{lrrrrr}
  \hline
ageclass & mean & median & sd & lower & upper \\ 
  \hline
[0,6) & 0.009 & 0.000 & 0.034 & 0.000 & 0.071 \\\relax 
  [6,12) & 0.018 & 0.000 & 0.086 & 0.000 & 0.300 \\\relax 
  [12,18) & 0.036 & 0.000 & 0.103 & 0.000 & 0.295 \\\relax 
  [18,30) & 1.000 & 0.639 & 1.136 & 0.000 & 3.763 \\\relax 
  [30,40) & 0.108 & 0.012 & 0.232 & 0.000 & 0.897 \\\relax 
  [40,50) & 5.047 & 5.173 & 1.413 & 1.977 & 7.763 \\\relax 
  [50,60) & 4.143 & 4.061 & 1.254 & 1.574 & 6.643 \\\relax 
  [60,70) & 4.625 & 4.608 & 0.839 & 3.281 & 6.540 \\\relax 
  [70,80) & 2.149 & 2.073 & 1.013 & 0.353 & 4.318 \\ 
  80+ & 0.394 & 0.154 & 0.556 & 0.000 & 1.739 \\ 
   \hline
\end{tabular}
\caption{Waves 12 13} 
\end{subtable}
\qquad
\begin{subtable}{0.48\textwidth}
\centering
\begin{tabular}{lrrrrr}
  \hline
ageclass & mean & median & sd & lower & upper \\ 
  \hline
[0,6) & 0.340 & 0.267 & 0.330 & 0.000 & 1.124 \\\relax 
  [6,12) & 0.117 & 0.023 & 0.214 & 0.000 & 0.711 \\\relax 
  [12,18) & 0.180 & 0.029 & 0.271 & 0.000 & 0.943 \\\relax 
  [18,30) & 0.445 & 0.147 & 0.632 & 0.000 & 2.034 \\\relax 
  [30,40) & 3.270 & 3.137 & 1.012 & 1.508 & 5.281 \\\relax 
  [40,50) & 2.581 & 2.458 & 1.152 & 0.335 & 5.005 \\\relax 
  [50,60) & 4.321 & 4.263 & 1.242 & 1.930 & 6.838 \\\relax 
  [60,70) & 4.689 & 4.919 & 1.575 & 1.894 & 7.435 \\\relax 
  [70,80) & 7.442 & 7.427 & 1.299 & 5.041 & 9.871 \\ 
  80+ & 1.252 & 1.047 & 1.024 & 0.000 & 3.903 \\ 
   \hline
\end{tabular}
\caption{Waves 14 15} 
\end{subtable}
\par\bigskip
\begin{subtable}{0.48\textwidth}
\centering
\begin{tabular}{lrrrrr}
  \hline
ageclass & mean & median & sd & lower & upper \\ 
  \hline
[0,6) & 0.453 & 0.359 & 0.474 & 0.000 & 1.516 \\\relax 
  [6,12) & 0.137 & 0.004 & 0.236 & 0.000 & 0.831 \\\relax 
  [12,18) & 0.350 & 0.172 & 0.468 & 0.000 & 1.580 \\\relax 
  [18,30) & 0.644 & 0.566 & 0.581 & 0.000 & 2.159 \\\relax 
  [30,40) & 1.541 & 1.399 & 1.168 & 0.000 & 3.898 \\\relax 
  [40,50) & 6.836 & 6.716 & 1.669 & 4.277 & 10.259 \\\relax 
  [50,60) & 0.436 & 0.114 & 0.613 & 0.000 & 2.063 \\\relax 
  [60,70) & 3.914 & 3.893 & 1.357 & 1.850 & 6.483 \\\relax 
  [70,80) & 1.468 & 1.447 & 0.988 & 0.016 & 3.418 \\ 
  80+ & 1.439 & 1.362 & 0.810 & 0.034 & 3.138 \\ 
   \hline
\end{tabular}
\caption{Waves 16 17} 
\end{subtable}
\qquad
\begin{subtable}{0.48\textwidth}
\centering
\begin{tabular}{lrrrrr}
  \hline
ageclass & mean & median & sd & lower & upper \\ 
  \hline
[0,6) & 0.128 & 0.042 & 0.180 & 0.000 & 0.618 \\\relax 
  [6,12) & 0.272 & 0.227 & 0.238 & 0.000 & 0.876 \\\relax 
  [12,18) & 1.008 & 0.938 & 0.523 & 0.178 & 2.250 \\\relax 
  [18,30) & 0.880 & 0.649 & 0.870 & 0.000 & 2.692 \\\relax 
  [30,40) & 0.864 & 0.774 & 0.601 & 0.003 & 2.178 \\\relax 
  [40,50) & 6.531 & 6.746 & 0.835 & 4.912 & 8.031 \\\relax 
  [50,60) & 3.524 & 3.749 & 0.945 & 1.635 & 5.077 \\\relax 
  [60,70) & 3.161 & 3.088 & 0.694 & 1.998 & 4.534 \\\relax 
  [70,80) & 0.600 & 0.447 & 0.542 & 0.000 & 1.683 \\ 
  80+ & 0.221 & 0.084 & 0.315 & 0.000 & 1.147 \\ 
   \hline
\end{tabular}
\caption{Waves 18 19} 
\end{subtable}
\par\bigskip
\begin{subtable}{0.48\textwidth}
\centering
\begin{tabular}{lrrrrr}
  \hline
ageclass & mean & median & sd & lower & upper \\ 
  \hline
[0,6) & 0.105 & 0.001 & 0.268 & 0.000 & 0.971 \\\relax 
  [6,12) & 0.031 & 0.000 & 0.112 & 0.000 & 0.263 \\\relax 
  [12,18) & 0.158 & 0.026 & 0.254 & 0.000 & 0.826 \\\relax 
  [18,30) & 1.088 & 0.833 & 0.897 & 0.000 & 2.888 \\\relax 
  [30,40) & 1.928 & 1.887 & 0.924 & 0.336 & 4.407 \\\relax 
  [40,50) & 9.128 & 9.524 & 1.643 & 5.465 & 12.172 \\\relax 
  [50,60) & 6.266 & 6.296 & 1.093 & 4.218 & 8.278 \\\relax 
  [60,70) & 0.338 & 0.178 & 0.386 & 0.000 & 1.188 \\\relax 
  [70,80) & 0.285 & 0.156 & 0.307 & 0.000 & 1.093 \\ 
  80+ & 0.113 & 0.029 & 0.263 & 0.000 & 0.675 \\ 
   \hline
\end{tabular}
\caption{Waves 20 21} 
\end{subtable}
\qquad
\begin{subtable}{0.48\textwidth}
\centering
\begin{tabular}{lrrrrr}
  \hline
ageclass & mean & median & sd & lower & upper \\ 
  \hline
[0,6) & 2.481 & 2.357 & 1.230 & 0.461 & 4.773 \\\relax 
  [6,12) & 0.365 & 0.219 & 0.450 & 0.000 & 1.416 \\\relax 
  [12,18) & 1.966 & 1.842 & 0.794 & 0.811 & 3.857 \\\relax 
  [18,30) & 7.831 & 7.097 & 2.411 & 4.653 & 12.346 \\\relax 
  [30,40) & 0.322 & 0.108 & 0.561 & 0.000 & 1.513 \\\relax 
  [40,50) & 1.436 & 1.264 & 1.076 & 0.000 & 4.204 \\\relax 
  [50,60) & 6.863 & 7.341 & 2.154 & 3.143 & 10.833 \\\relax 
  [60,70) & 0.174 & 0.079 & 0.305 & 0.000 & 0.773 \\\relax 
  [70,80) & 0.999 & 0.672 & 0.992 & 0.007 & 3.578 \\ 
  80+ & 0.214 & 0.104 & 0.274 & 0.000 & 1.010 \\ 
   \hline
\end{tabular}
\caption{Waves 22 23} 
\end{subtable}
\caption{Relative $q$-infectiousness with time evolution using assumption on susceptibility {\scriptsize$(1,1,1,1,1,1,1,1,1,1)$} corresponding to Figure \ref{inf6homo}.} 
\label{inf6homotable}
\end{table}

\clearpage
\subsection*{Estimation of $q$-infectiousness using heterogeneous susceptibility}

Method: Estimation of the $(h_j)$ relative $q$-infectiousness vector.\\
Assumption: heterogeneous susceptibility  $(a_i) = (0.4,0.39,0.38,0.79,0.86,0.8,0.82,0.88,0.74,0.74)$ taken from \cite{Davies:2020wt}.\\
Normalization method: Mean $q$-infectiousness among children age classes [0,6),  [6,12) and [12,18) is assumed constant among bootstraps and wave groups (if applicable). The mean of the first adult age class [18,30) is set to 1 for the first period.

\begin{table}[ht]
\scriptsize
\centering
\begin{tabular}{lrrrrr}
  \hline
ageclass & mean & median & sd & lower & upper \\ 
  \hline
  [0,6) & 0.346 & 0.034 & 0.461 & 0.000 & 1.483 \\\relax 
  [6,12) & 0.892 & 0.957 & 0.404 & 0.000 & 1.549 \\\relax 
  [12,18) & 1.310 & 1.326 & 0.231 & 0.870 & 1.710 \\\relax 
  [18,30) & 1.000 & 0.967 & 0.544 & 0.000 & 2.051 \\\relax 
  [30,40) & 0.645 & 0.375 & 0.811 & 0.000 & 2.512 \\\relax 
  [40,50) & 3.783 & 3.541 & 1.586 & 1.672 & 7.443 \\\relax 
  [50,60) & 1.320 & 1.257 & 0.741 & 0.000 & 2.705 \\\relax 
  [60,70) & 0.266 & 0.000 & 0.445 & 0.000 & 1.586 \\\relax 
  [70,80) & 1.277 & 1.155 & 0.903 & 0.000 & 3.362 \\ 
  80+ & 0.099 & 0.000 & 0.279 & 0.000 & 0.950 \\ 
   \hline
\end{tabular}
\caption{Relative $q$-infectiousness using assumption on susceptibility {\scriptsize$(0.4,0.39,0.38,0.79,0.86,0.8,0.82,0.88,0.74,0.74)$} corresponding to Figure \ref{inf1ass}.} 
\label{inf1asstable}
\end{table}

\begin{table}[ht]
\scriptsize
\begin{subtable}{0.48\textwidth}
\centering
\begin{tabular}{lrrrrr}
  \hline
ageclass & mean & median & sd & lower & upper \\ 
  \hline
[0,6) & 0.204 & 0.013 & 0.363 & 0.000 & 1.319 \\\relax 
  [6,12) & 0.483 & 0.454 & 0.383 & 0.000 & 1.402 \\\relax 
  [12,18) & 1.061 & 1.070 & 0.443 & 0.001 & 1.748 \\\relax 
  [18,30) & 1.000 & 0.625 & 1.259 & 0.000 & 3.990 \\\relax 
  [30,40) & 1.625 & 0.052 & 12.906 & 0.000 & 4.577 \\\relax 
  [40,50) & 9.694 & 3.895 & 49.732 & 0.345 & 25.974 \\\relax 
  [50,60) & 5.792 & 2.434 & 31.815 & 0.000 & 17.774 \\\relax 
  [60,70) & 0.398 & 0.063 & 0.802 & 0.000 & 2.223 \\\relax 
  [70,80) & 2.973 & 2.087 & 4.053 & 0.018 & 10.597 \\ 
  80+ & 0.189 & 0.000 & 0.505 & 0.000 & 1.891 \\ 
   \hline
\end{tabular}
\caption{Waves 12 13} 
\end{subtable}
\qquad
\begin{subtable}{0.48\textwidth}
\centering
\begin{tabular}{lrrrrr}
  \hline
ageclass & mean & median & sd & lower & upper \\ 
  \hline
[0,6) & 0.560 & 0.547 & 0.453 & 0.000 & 1.535 \\\relax 
  [6,12) & 0.464 & 0.404 & 0.436 & 0.000 & 1.370 \\\relax 
  [12,18) & 0.724 & 0.760 & 0.278 & 0.005 & 1.134 \\\relax 
  [18,30) & 0.484 & 0.287 & 0.554 & 0.000 & 1.814 \\\relax 
  [30,40) & 0.818 & 0.700 & 0.695 & 0.000 & 2.559 \\\relax 
  [40,50) & 1.404 & 1.172 & 1.127 & 0.000 & 4.761 \\\relax 
  [50,60) & 0.838 & 0.739 & 0.711 & 0.000 & 2.570 \\\relax 
  [60,70) & 1.737 & 1.647 & 1.033 & 0.000 & 4.353 \\\relax 
  [70,80) & 4.463 & 4.026 & 1.995 & 2.142 & 9.635 \\ 
  80+ & 0.549 & 0.501 & 0.550 & 0.000 & 1.707 \\ 
   \hline
\end{tabular}
\caption{Waves 14 15} 
\end{subtable}
\par\bigskip
\begin{subtable}{0.48\textwidth}
\centering
\begin{tabular}{lrrrrr}
  \hline
ageclass & mean & median & sd & lower & upper \\ 
  \hline
[0,6) & 0.433 & 0.425 & 0.412 & 0.000 & 1.167 \\\relax 
  [6,12) & 0.436 & 0.388 & 0.398 & 0.000 & 1.405 \\\relax 
  [12,18) & 0.879 & 0.897 & 0.267 & 0.161 & 1.322 \\\relax 
  [18,30) & 0.246 & 0.153 & 0.311 & 0.000 & 1.093 \\\relax 
  [30,40) & 0.479 & 0.212 & 0.661 & 0.000 & 2.002 \\\relax 
  [40,50) & 2.615 & 2.145 & 1.843 & 0.488 & 6.913 \\\relax 
  [50,60) & 0.133 & 0.000 & 0.341 & 0.000 & 1.184 \\\relax 
  [60,70) & 0.495 & 0.400 & 0.432 & 0.000 & 1.497 \\\relax 
  [70,80) & 1.725 & 1.432 & 1.105 & 0.303 & 4.271 \\ 
  80+ & 0.674 & 0.501 & 0.764 & 0.000 & 2.751 \\ 
   \hline
\end{tabular}
\caption{Waves 16 17} 
\end{subtable}
\qquad
\begin{subtable}{0.48\textwidth}
\centering
\begin{tabular}{lrrrrr}
  \hline
ageclass & mean & median & sd & lower & upper \\ 
  \hline
[0,6) & 0.189 & 0.146 & 0.177 & 0.000 & 0.556 \\\relax
  [6,12) & 0.480 & 0.484 & 0.147 & 0.196 & 0.740 \\\relax 
  [12,18) & 1.079 & 1.090 & 0.142 & 0.806 & 1.342 \\\relax 
  [18,30) & 0.333 & 0.320 & 0.230 & 0.000 & 0.806 \\\relax 
  [30,40) & 0.181 & 0.139 & 0.178 & 0.000 & 0.548 \\\relax 
  [40,50) & 1.536 & 1.418 & 0.703 & 0.520 & 3.044 \\\relax 
  [50,60) & 0.212 & 0.143 & 0.235 & 0.000 & 0.820 \\\relax 
  [60,70) & 0.120 & 0.083 & 0.127 & 0.000 & 0.384 \\\relax 
  [70,80) & 0.469 & 0.432 & 0.328 & 0.001 & 1.227 \\ 
  80+ & 0.089 & 0.021 & 0.115 & 0.000 & 0.354 \\ 
   \hline
\end{tabular}
\caption{Waves 18 19} 
\end{subtable}
\par\bigskip
\begin{subtable}{0.48\textwidth}
\centering
\begin{tabular}{lrrrrr}
  \hline
ageclass & mean & median & sd & lower & upper \\ 
  \hline
[0,6) & 0.255 & 0.038 & 0.367 & 0.000 & 1.115 \\\relax 
  [6,12) & 0.646 & 0.662 & 0.376 & 0.018 & 1.484 \\\relax 
  [12,18) & 0.847 & 0.867 & 0.335 & 0.121 & 1.433 \\\relax 
  [18,30) & 0.682 & 0.607 & 0.546 & 0.000 & 2.017 \\\relax 
  [30,40) & 0.790 & 0.709 & 0.631 & 0.000 & 2.279 \\\relax 
  [40,50) & 4.716 & 4.233 & 2.591 & 2.017 & 10.348 \\\relax 
  [50,60) & 0.112 & 0.002 & 0.262 & 0.000 & 0.999 \\\relax 
  [60,70) & 0.049 & 0.003 & 0.149 & 0.000 & 0.461 \\\relax 
  [70,80) & 0.072 & 0.004 & 0.196 & 0.000 & 0.585 \\ 
  80+ & 0.111 & 0.004 & 0.248 & 0.000 & 0.851 \\ 
   \hline
\end{tabular}
\caption{Waves 20 21} 
\end{subtable}
\qquad
\begin{subtable}{0.48\textwidth}
\centering
\begin{tabular}{lrrrrr}
  \hline
ageclass & mean & median & sd & lower & upper \\ 
  \hline
[0,6) & 0.794 & 0.823 & 0.321 & 0.013 & 1.352 \\\relax 
  [6,12) & 0.263 & 0.208 & 0.247 & 0.000 & 0.799 \\\relax 
  [12,18) & 0.692 & 0.689 & 0.189 & 0.339 & 1.044 \\\relax 
  [18,30) & 0.671 & 0.644 & 0.274 & 0.210 & 1.346 \\\relax 
  [30,40) & 0.039 & 0.000 & 0.109 & 0.000 & 0.323 \\\relax 
  [40,50) & 0.496 & 0.436 & 0.404 & 0.000 & 1.327 \\\relax 
  [50,60) & 0.399 & 0.391 & 0.270 & 0.000 & 0.918 \\\relax 
  [60,70) & 0.014 & 0.000 & 0.042 & 0.000 & 0.123 \\\relax 
  [70,80) & 0.104 & 0.053 & 0.144 & 0.000 & 0.507 \\ 
  80+ & 0.016 & 0.004 & 0.033 & 0.000 & 0.105 \\ 
   \hline
\end{tabular}
\caption{Waves 22 23} 
\end{subtable}
\caption{Relative $q$-infectiousness with time evolution using assumption on susceptibility {\scriptsize$(0.4,0.39,0.38,0.79,0.86,0.8,0.82,0.88,0.74,0.74)$} corresponding to Figure \ref{inf6ass}.} 
\label{inf6asstable}
\end{table}

\end{document}